\documentclass[10pt]{article}
\usepackage{amsmath}
\usepackage{amsthm}
\usepackage{amsfonts}
\usepackage{amssymb}
\usepackage{url}
\usepackage{hyperref}
\usepackage{eucal}
\usepackage{mathrsfs}
\usepackage{graphicx,graphics}
\usepackage[T1]{fontenc}    
\usepackage{url}

\newcommand\labfig[1] {\label{fig:#1}}
\newcommand\sect[1] {\ref{sect:#1}}
\newcommand\labsect[1] {\label{sect:#1}}

\usepackage{graphicx}
\newcommand{\bfm}[1]{\mbox{\boldmath ${#1}$}}
\newcommand{\nonum}{\nonumber \\}
\newcommand\eq[1] {(\ref{#1})} 
\newcommand{\beqa}{\begin{eqnarray}}
\newcommand{\eeqa}[1]{\label{#1}\end{eqnarray}}
\newcommand{\beq}{\begin{equation}}
\newcommand{\eeq}[1]{\label{#1}\end{equation}}
\newcommand{\R}{\mathbb{R}}

\newcommand{\Grad}{\nabla}
\newcommand{\Div}{\nabla \cdot}

\newcommand{\Real}{\mathop{\rm Re}\nolimits}
\newcommand{\Imag}{\mathop{\rm Im}\nolimits}
\newcommand{\Tr}{\mathop{\rm Tr}\nolimits}

\newcommand{\lang}{\langle}
\newcommand{\rang}{\rangle}
\newcommand{\Md}{\partial}

\newcommand{\Ga}{\alpha}

\newcommand{\Ge}{\epsilon}

\newcommand{\Gc}{\chi}

\newcommand{\Gk}{\kappa}

\newcommand{\Gl}{\lambda}

\newcommand{\Gm}{\mu}

\newcommand{\Gt}{\theta}

\newcommand{\Gr}{\rho}
\newcommand{\Gvr}{\varrho}

\newcommand{\Go}{\omega}

\newcommand{\GP}{\Pi}

\newcommand{\GO}{\Omega}

\newcommand{\BGa}{\bfm\alpha}

\newcommand{\BGe}{\bfm\epsilon}

\newcommand{\BGr}{\bfm\rho}

\newcommand{\BGs}{\bfm\sigma}

\newcommand{\BGL}{\bfm\Lambda}
\newcommand{\BGP}{\bfm\Pi}

\newcommand{\CA}{{\cal A}}

\newcommand{\CE}{{\cal E}}
\newcommand{\CF}{{\cal F}}
\newcommand{\CG}{{\cal G}}
\newcommand{\CH}{{\cal H}}

\newcommand{\CJ}{{\cal J}}
\newcommand{\CK}{{\cal K}}

\newcommand{\CP}{{\cal P}}
\newcommand{\CQ}{{\cal Q}}

\newcommand{\CS}{{\cal S}}

\newcommand{\CV}{{\cal V}}

\newcommand{\BCL}{{\bfm{\cal L}}}

\newcommand{\BCY}{{\bfm{\cal Y}}}

\newcommand{\bpm}{\begin{pmatrix}}
\newcommand{\epm}{\end{pmatrix}}

\def\b0{\bf 0}

\def\Bb{{\bf b}}

\def\Bd{{\bf d}}
\def\Be{{\bf e}}

\def\Bk{{\bf k}}

\def\Bm{{\bf m}}
\def\Bn{{\bf n}}

\def\Bp{{\bf p}}
\def\Bq{{\bf q}}
\def\Br{{\bf r}}
\def\Bs{{\bf s}}
\def\Bt{{\bf t}}
\def\Bu{{\bf u}}
\def\Bv{{\bf v}}

\def\Bx{{\bf x}}

 \def\BE{{\bf E}}
\def\BF{{\bf F}}
\def\BG{{\bf G}}
\def\BH{{\bf H}}
\def\BI{{\bf I}}
\def\BJ{{\bf J}}
\def\BK{{\bf K}}
\def\BL{{\bf L}}
\def\BM{{\bf M}}

\def\BP{{\bf P}}

\def\BR{{\bf R}}

\def\BY{{\bf Y}}
\def\BZ{{\bf Z}}

\title{Bounds on complex polarizabilities and a new perspective on scattering by a lossy inclusion}
\date{}
\begin{document}
\maketitle
\vskip -.5cm
\centerline{\large
Graeme W. Milton \footnote{Department of Mathematics, University of Utah, USA -- milton@math.utah.edu,},}
\vskip 1.cm
\begin{abstract}
Here we obtain explicit formulae for bounds on the complex electrical polarizability at a given frequency 
of an inclusion with known volume that 
follow directly from the quasistatic bounds of Bergman and Milton on the effective complex dielectric constant
of a two-phase medium. We also describe how analogous bounds on the orientationally averaged bulk and shear
polarizabilities at a given frequency can be obtained from bounds on the effective complex bulk and shear
moduli of a two-phase medium obtained by Milton, Gibiansky and Berryman, using the quasistatic variational
principles of Cherkaev and Gibiansky. We also show how the polarizability problem and the acoustic scattering problem
can both be reformulated in an abstract setting as ``$Y$-problems''. In the acoustic scattering context, to avoid explicit
introduction of the Sommerfeld radiation condition, we introduce auxilliary fields at infinity and an appropriate
``constitutive law'' there, which forces the Sommerfeld radiation condition to hold. As a consequence we obtain minimization variational
principles for acoustic scattering that can be used to obtain bounds on the complex backwards scattering amplitude.
Some explicit elementary bounds are given.

\end{abstract}
\section{Introduction}
\setcounter{equation}{0}
Here we consider scattering of waves by lossy inclusions. By lossy we mean that the inclusion absorbs
 energy. If the wavelength inside and outside the inclusion, and attenuation lengths inside the inclusion, are very long compared to the 
diameter of the inclusion then one may use a quasistatic approximation, where one uses the usual static equations but with complex 
valued fields and complex valued material moduli. At fixed frequency $\Go$ the physical fields in the neighborhood of the inclusion are 
obtained by multiplying these complex fields by $e^{-i\Go t}$ and then taking the real part. The leading correction to the field at
long distances from the inclusion, long compared to the diameter but short compared to the relevant wavelengths or attenuation lengths,
is the dipolar part and the relation between it and the incident field is governed by the polarizability of the inclusion. 

In the context of the dielectric problem, a dilute array of scatterers each with polarizability matrix $\BGa$, but randomly orientated
so the average polarizability is $(\Tr\BGa/3)\BI$, has an effective dielectric constant well known to be 
\beq \Ge_*\approx 1+p\Tr(\BGa)/(3|\GO|), \eeq{0.01}
where $\GO$ is the volume of the inclusion, and $p$ is the volume occupied by the inclusion phase in the array.
Thus the low volume fraction limit of the microstructure independent Bergman-Milton bounds 
\cite{Bergman:1980:ESM,Milton:1980:BCD,Milton:1981:BCP,Milton:1981:BTO}
on the complex dielectric constant $\Ge_*$ 
of a isotropic two-phase composite immediately give one bounds on the complex average polarizability. 
In this way, bounds on complex polarizabilities were obtained as long ago as 1979 \cite{Milton:1979:TST},
although it was not until 1981 that the results were published (see figure 3 in \cite{Milton:1981:TPA},
reproduced here in figure 1(b)).
The Bergman-Milton bounds were obtained via the analytic approach -- using the analytic properties of the
effective dielectric constant as a function of the component dielectric constants. From a wider perspective
the bounds are related to bounds on Stieltjes and Herglotz functions, and to the Nevanlinna-Pick interpolation
problem on which there is a huge literature. In the case where the bounds on the complex dielectric constant $\Ge_*$ are sharp, 
such as in two-dimensions \cite{Milton:1980:BCD, Milton:1981:BCP, Milton:1981:BTO}, then the corresponding bounds on the 
complex polarizabilities are also, at least asymptotically, sharp. We mention that analytic representations,
similar to those obtained for the effective moduli of composites \cite{Bergman:1978:DCC,Milton:1981:BCP,Golden:1983:BEP,Kantor:1982:ERN,Milton:2002:TOC},
have also been obtained for the polarizability tensor \cite{Fuchs:1975:TOP,Fuchs:1976:SRP,Cassier:2017:BHF}, and for
electromagnetic scattering \cite{Bergman:1980:TRE,Bergman:2016:EEF}.
\begin{figure}[ht]
\centering
\includegraphics[width=0.75\textwidth]{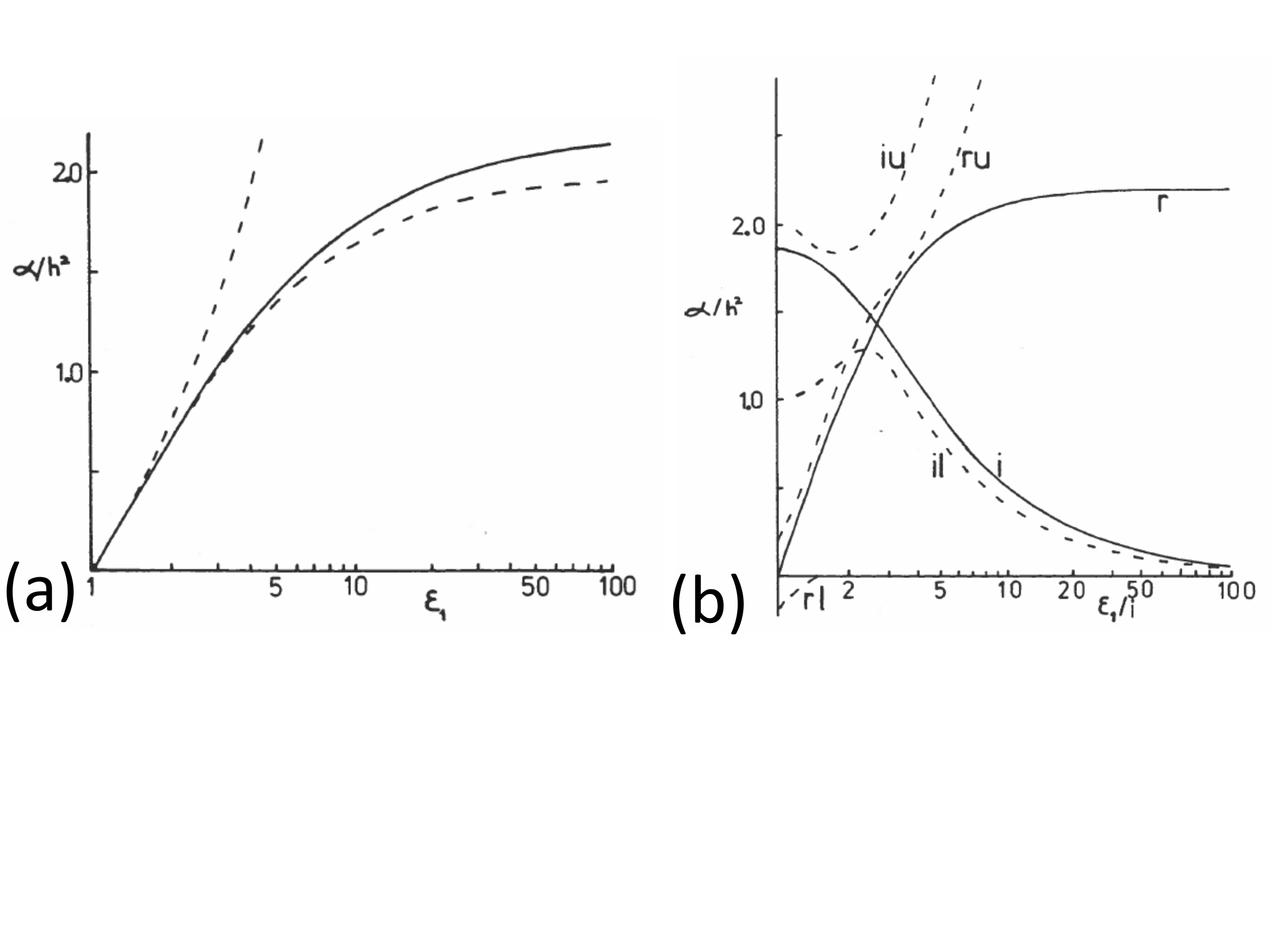}
\vspace{-1in}
\caption{The dashed lines show bounds on $\Ga/h^2$ representing the orientationaly averaged real and complex polarizability per unit volume of an arbitarily shaped two-dimensional
inclusion. In (a) the dielectric constant $\Ge_1$ of the inclusion is real, while in (b) it is purely imaginary. The surrounding medium has a dielectric constant of unity.
In (b) ``ru'' and ``rl'' denote the upper and lower
bounds on the real part of $\Ga/h^2$, while ``iu'' and ``il'' denote the upper and lower bounds on the imaginary part of $\Ga/h^2$. The solid lines are the numerical results
for a square shaped inclusion. The bounds in (a) are asymptotically attained in the cases of the solid circular cylinder and the thin cylindrical shell.
The figures are reproductions, with permission of Springer, of figures 2 and 3 in \protect\cite{Milton:1981:TPA}.}
\labfig{fig:polar}
\label{fig:1}
\end{figure}

Recently there has been a resurgence of interest in such bounds on the complex polarizability, or at least the imaginary part which
governs the absorption. This is fed by the realization that such bounds are helpful to determine the absorption of radiation
of a cloud of dispersed sub-wavelength sized metal particles that may be useful for smoke screens \cite{Miller:2014:FLE}. 
The authors of \cite{Miller:2014:FLE} apparently  did not realize that bounds on the complex quaistatic polarizability are in fact a simple corollary of 
those on the complex dielectric constant of periodic two-phase composites in the small volume fraction limit. 

The bounds on the complex dielectric constant have also been obtained using the variational principles of Cherkaev and
Gibiansky \cite{Cherkaev:1994:VPC}. In fact, for viscoelastic problems at fixed frequency where one is interested in bounding
the complex effectve elasticity tensor it seems that the variational approach is more suitable than the analytic approach
\cite{Gibiansky:1993:EVM, Milton:1997:EVM, Gibiansky:1999:EVM}.
Both the variational approach and the the analytic approach have been extended to viscoelastic problems in the time
domain, by Carini and Mattei \cite{Carini:2015:VFL} and Mattei and Milton \cite{Mattei:2016:BRV} respectively.
In this connection, for obtaining bounds on the viscoelastic response at a given time, it seems that the
analytic approach is the most suitable method. 

Most interesting has been the recent breakthrough result of Miller et.al. \cite{Miller:2015:FLO} where through astoundingly simple arguments
they obtain inclusion shape independent bounds on the scattered power, absorbed power, and their sum (known as the extinction) in 
terms of the material moduli, frequency, and amplitude of the incident plane wave. Most significantly they do not assume that
the inclusion is small compared to the wavelength: they use the full time-harmonic Maxwell equations rather than just the quasistatic approximation.

Thus one wonders if there are some variational minimization principles that apply to scattering by an inclusion. Here we will see
that indeed there are such variational minimization principles. However with a choice of trial fields, they do not
provide a bound on the extinction, or equivalently the forward scattering amplitude, but rather surprisingly provide a bound on the
{\it backward scattering amplitude}. Thus it seems that these variational principles do not allow one to recover
the extinction bounds of  Miller et.al. \cite{Miller:2015:FLO}. Our approach to obtaining variational principles follows
that in chapter 12 of \cite{Milton:2002:TOC}: since the equations are linear, the variational principles should be quadratic
and obtained by expanding a positive semidefinite quadratic form of the difference between the actual fields and the trial fields,
where those terms in the expansion that involve products of the actual field and the trial field need to be integrated by parts
(or equivalently evaluated using the orthogonality properties of the relevant subspaces of fields). 

We mention that minimization principles have been obtained by Milton, Seppecher and Bouchitt{\'e} \cite{Milton:2009:MVP}
and Milton and Willis \cite{Milton:2010:MVP} for the full time harmonic acoustic equations, Maxwell's equations,
and elastodynamic equations, in bodies of finite extent containing inhomogeneous lossy media. This advance was made possible by the key realization 
that these equations can all be suitably manipulated into a form where it is easy to see that one can directly apply the transformation techniques of Cherkaev and
Gibiansky \cite{Cherkaev:1994:VPC} to obtain minimization variational principles. While is not immediately clear how to extend
these variational principles to scattering, this is in fact what we will ultimately succeed in doing. For simplicity, we confine our
attention to the acoustic problem: electromagnetic and elastodynamic scattering will be considered elsewhere. 

We will see that problems of determining polarizabilty tensors and solving scattering by an inclusion can be naturally formulated in an abstract
setting as ``$Y$-problems''. For an introduction to ``$Y$-problems'' and their significance, see Chapters 19, 20, and 29 in
the book \cite{Milton:2002:TOC}, as well as Sections 23.6, 23.7, and 24.10 therein, and also Chapters 1,2,7, 9, and 10 in the book \cite{Milton:2016:ETC}. Briefly, ``Y-tensors'', and the associated fractional linear transformations linking effective tensors and ``Y-tensors'' first appeared in bounds on the effective moduli of composites, in 
formulae for effective medium approximations, and in continued fractions for the effective tensor 
\cite{Milton:1981:BEE,Berryman:1982:EMT,Milton:1985:TCC,Milton:1987:MCEa,Cherkaev:1992:ECB}. The continued fractions were connected with a 
hierarchical spitting of the relevant
Hilbert space, known as the field equation recursion method, in which ``$Y$-problems'' make a natural  appearance at successive
stages of the procedure \cite{Milton:1987:MCEa,Milton:1991:FER}.
In the first stage of the procedure, for a two-phase periodic composite, the tensor $\BY_*$ was found to have a direct physical meaning, relating
the phase averages of the fluctuating components of the fields \cite{Gibiansky:1993:EVM}. For example, in a dielectric problem with a periodic dielectric constant, 
a periodic displacement field $\Bd(\Bx)$ and periodic electric field $\Be(\Bx)$, one has
\beq \lang\chi_i(\Bd-\lang\Bd\rang)\rang=-\BY_*\lang\chi_i(\Be-\lang\Be\rang)\rang, \eeq{0.0}
where $\chi_i(\Bx)$, $i=1,2$, is the indicator function taking the value $1$ in phase $i$, and $0$ in the other phase, and the angular brackets $\lang\cdot\rang$ denote
a volume average over the unit cell of periodicity.

The setting of a $Y$-problem is a Hilbert space, or finite dimensional vector space, $\CK$ that has the decomposition
\beq \CK=\CE\oplus\CJ=\CV\oplus\CH, \eeq{0.1}
where the spaces $\CE$ and $\CJ$ are orthogonal complements, as are the spaces $\CV$ and $\CH$. Given a linear operator $\BL$
mapping $\CH$ to $\CH$, the $Y$-problem is to find for each given element $\BE_1$ of $\CV$ the associated fields
\beq \BE_2,\BJ_2\in\CH,\quad \BJ_1\in\CV,\quad {\rm such~that~}\BE=\BE_1+\BE_2\in\CE,\quad \BJ=\BJ_1+\BJ_2\in\CJ, \quad \BJ_2=\BL\BE_2.
\eeq{0.2}
Note that because $\CV$ and $\CH$ are orthogonal and span $\CK$ any field, or vector, $\BK\in\CK$ can be split into $\BK=\BK_1+\BK_2$,
where $\BK_1\in\CV$ and $\BK_2\in\CH$. Assuming that these fields are uniquely determined for each $\BE_1\in\CV$, $\BJ_1$ must be linearly dependent on $\BE_1$ and this
linear relation,
\beq \BJ_1=-\BY_*\BE_1, \eeq{0.3}
defines the associated operator $\BY_*$, which maps $\CV$ to $\CV$, or to a subspace of $\CV$. The meaning of the spaces $\CK$, $\CE$, $\CJ$, $\CV$ and $\CH$
will of course depend on the problem under consideration and many examples can be given. In the context of a two-phase dielectric periodic composite, as in \eq{0.0},
$\CK$ is the space of square integrable periodic fields with zero average over the unit cell, $\CE$ is the space of gradients of periodic potentials, $\CJ$
are those fields in $\CK$ that have zero divergence, $\CV$ are those fields in $\CK$ that are constant in each phase, $\CH$ are those fields whose average
over each phase is zero.  Another concrete example is an electrical network
comprised of a network of $m$, possibly complex, impedances on one side of the circuit board, and a network of $b$ batteries or oscillating power souces
on the other side of the circuit board, with
the two networks being connected by terminal nodes drilled through the circuit board. Fields in $\CK$ are then $m+b$ dimensional
vectors whose elements represent the field components in the impedances or batteries. Fields in $\CE$ are potential drops, while fields
in $\CJ$ represent currents satisfying the condition that the net flux of current in or out of any node is zero. The subspace $\CE$ 
can also be seen as the column space of the incidence matrix $\BM$ of the entire network, and the subspace $\CJ$ as the null-space of $\BM^T$, thus
accounting for the orthogonality of these subspaces. Fields in $\CV$ have elements
which are non-zero only in the batteries, while fields in $\CH$ have elements which are non-zero only in the impedances. The matrix $\BL$ is then
diagonal with elements representing the individual impedance values. The tensor $\BY_*$ measures the response of the batteries. The orthogonality
of $\BE$ and $\BJ$ coupled with the orthogonality of implies $-\BJ_1\cdot\BE_1=\BJ_2\cdot\BE_2$ or equivalently that
\beq \BE_1\BY_*\BE_1=\BE_2\cdot\BL\BE_2. \eeq{0.4}
So if $\BL$ is real and positive semidefinite (as is the case when the impedances are resistors), then \eq{0.4} is a restatement of the fact that the net power
provided by the batteries is equal to the net power consumed by the impedances. It also implies $\BY_*$ is positive semidefinite, which is why
the minus sign is introduced in the definition \eq{0.3}. For more details see Chapter 19 in \cite{Milton:2002:TOC}. Interestingly, one can
perform algebraic operations on  $Y$-problems, in the same way that one can perform algebraic operations such as addition, multiplication and substitution
with electrical circuits, and moreover if one removes the orthogonality constraints on the subspaces these operations can be extended to include subtraction
and division: one has a complete algebra (see Chapter 7 in \cite{Milton:2016:ETC}). 

The advantage of recognizing 
that determining polarizabilty tensors and solving scattering by an inclusion are both ``$Y$-problems'' is that one can
more or less immediately write down variational minimization principles, even when the moduli are complex, and also
one can deduce important analytic properties of $\BY_*$ as a function of the component moduli. Both the variational principles
and the analytic properties can lead to bounds on $\BY_*$, and thus to bounds on the polarizabilty tensor, or on the
scattering amplitudes. 

\section{Formulating the problem of determining the polarizability tensor as a $Y$-problem}
\setcounter{equation}{0}
The purpose of this section is twofold: first, to introduce $Y$-problems in a simple setting quite close to that of acoustic scattering,
namely the dielectric problem, in quasistatics, of determining the complex polarizability tensor of a lossy inclusion in a 
three-dimensional infinite homogeneous dielectric medium; and second to review the accompanying standard 
analysis as it will have direct parallels in the context of acoustic scattering. In a two-phase periodic composite the
simplest associated $Y$-problem is obtained by stripping the constant fields from the underlying equations (see,
for example, \cite{Gibiansky:1993:EVM} and section 19.1 in \cite{Milton:2002:TOC}). Similarly for the polarizabilty problem,
the associated $Y$-problem is obtained by stripping the constant applied incident fields from the underlying equations.

The permittivity $\BGe(\Bx)$ is $\BGe_1$ inside the inclusion and
$\BGe_0$ outside:
\beq \BGe(\Bx)=\BGe_0+(\BGe_1-\BGe_0)\Gc(\Bx), \eeq{1.7}
where $\Gc(\Bx)$ is the indicator function taking the value 1 in the inclusion and 0 outside. Let $\CK$ denote the Hilbert space of square integrable $3$-component
vector fields. Then the constitutive law takes the form
\beq \underbrace{\Bd_0+\Bd^s(\Bx)}_{\Bd(\Bx)}=\BGe(\Bx)(\underbrace{\Be_0+\Be^s(\Bx)}_{\Be(\Bx)}), \eeq{1.5}
where $\Bd_0$ and $\Be_0$ are constant fields, with $\Bd_0=\BGe_0\Be_0$,  while 
\beq \Be^s\in\CE,\quad \Bd^s\in\CJ,\eeq{1.6}
in which $\CE$ is the space of fields in $\CK$ that 
have zero curl, while $\CJ$ is the space of fields in $\CK$ that 
have zero divergence. For simplicity the dielectric tensor outside is assumed to be isotropic, of the
form $\BGe_0=\Ge_0\BI$, where $\Ge_0$ is a positive scalar. The electric potential $V(\Bx)$ outside any
sphere containing the inclusion has an expansion in spherical harmonics \cite{Jackson:1975:CEa}, the leading term of which is
\beq V^s(\Bx)=\Bb\cdot\Bx/(4\pi\Ge_0 r^3)  +\cdots{},
\eeq{8a-1}
and the associated electric field $\Be^s(\Bx)=-\Grad V^s(\Bx)$ is
\beq \Be^s=-\Grad V^s=-\Bb/(4\pi\Ge_0 r^3) +3\Bx(\Bb\cdot\Bx)/(4\pi\Ge_0  r^5)
 +\cdots{}.
\eeq{8a-2}
So we see that at large distances the dominant correction to the uniform field
comes from terms involving the vector $\Bb$; this vector is known as
the induced dipole moment.%
\index{dipole moment}
The factor of $4\pi\Ge_0$ has been introduced into
the above expansions so that $\Bb$ has a physical interpretation when
inclusion is in free space and $\Ge_0$ represents the dielectric constant
(or, more precisely, the electrical permittivity) of free space. As we will
see shortly, $\Bb$ can then be identified with the first moment of the
induced charge density.

Since the equations for the fields are linear, there must be
a linear relation between the induced dipole moment $\Bb$ and the applied
field $\Be_0$. This linear
relation,
\beq \Bb=\BGa\Be_0, \eeq{1.7a}
defines the polarizability tensor%
\index{polarizability!tensor}
$\BGa$ of the inclusion. This tensor
has also been called the P\'olya-Szeg{\H{o}} matrix; see \cite{Polya:1951:IIM}
and \cite{Movchan:1995:MMS}.

For a fixed
applied field $\Be_0$ the vector $\Bb$ is determined by the integral
of the polarization field,
\beq \Bp(\Bx)=(\Ge(\Bx)-\Ge_0)\Be(\Bx)=\Bd(\Bx)-\Ge_0\Be(\Bx)=\Bd_0+\Bd^s(\Bx)-\Ge_0[\Be_0+\Be^s(\Bx)]
=\Bd^s(\Bx)-\Ge_0\Be^s(\Bx),
\eeq{1.8}
over the volume of the inclusion. To see this we follow, for example, 
the argument given in Section 10.1 of \cite{Milton:2002:TOC}. Consider a ball $B_{r_0}$ of
very large radius $r$ containing the inclusion. Since the polarization
field is zero outside the inclusion, we can equate the integral of
the polarization field over the inclusion with the integral of
the polarization field over the ball $B_{r_0}$. Since the displacement field
$\Bd(\Bx)$ has zero divergence, and since $-\Be^s(\Bx)$ is the
gradient of the electrical potential $V^s(\Bx)$, it follows that
for any vector $\Bm$
\beqa \int_{B_{r_0}}\Bm\cdot\Bp(\Bx)\,d\Bx
&  = &\int_{B_{r_0}}\Bd^s(\Bx)\cdot\Grad(\Bm\cdot\Bx)\,d\Bx+\Ge_0\Bm\cdot\int_{B_{r_0}}\Grad V^s(\Bx)\,d\Bx
(\Bx)
\nonum
& = &\int_{\partial B_{r_0}}(\Bm\cdot\Bx)\Bd^s(\Bx)\cdot\Bn
+\Ge_0 V^s(\Bx)\Bm\cdot\Bn \,dS
\nonum
& = &\Ge_0\Bm\cdot\int_{\partial B_{r_0}}V^s(\Bx)\Bn-\Bx(\Grad V^s(\Bx)\cdot\Bn)\,dS,
\eeqa{1.9}
where $\Bn=\Bx/|\Bx|$ is the outward normal to the surface $\partial B$ of the
ball $ B$. When the radius $r$ of the ball $ B$ is sufficiently
large we can use the asymptotic formulas \eq{8a-1} and \eq{8a-2} to estimate
these integrals,
\beqa \int_{\partial B_{r_0}}\Bx(\Grad V^s(\Bx)\cdot\Bn)\,dS,
& \approx &-\int_{\partial B_{r_0}}2\Bx(\Bb\cdot\Bx)/(4\pi\Ge_0 r_0^4)\,dS
=\frac{2}{3\Ge_0}\Bb, \nonum
\int_{\partial B_{r_0}} V(\Bx)\Bn\,dS,
& \approx & \int_{\partial B_{r_0}}\Bx(\Bb\cdot\Bx)/(4\pi\Ge_0 r_0^4)\,dS
=\frac{1}{3\Ge_0}\Bb,
\eeqa{1.10}
with these approximations becoming increasingly accurate as
the radius $r$ of the ball $B_{r_0}$ approaches infinity.
By subtracting these expressions and taking the limit as $r$ approaches
infinity we see that
\beq \int_{B_{r_0}}\Bp(\Bx)\,d\Bx=\Bb.
\eeq{8a-7}

Now we define $\CV$ to consist of all fields of the form $\Gc(\Bx)\Bv$ where $\Bv$ is a constant vector, i.e., which are constant
in the inclusion and zero outside, and we define $\CH$ as the orthogonal complement of $\CV$ in the subspace $\CK$, i.e.,
those fields in $\CK$ that have zero average value over the inclusion. Then we rewrite \eq{1.5} as
\beq \Bd^s(\Bx)=\BGe(\Bx)\Be^s(\Bx)+(\BGe_1-\BGe_0)\Gc(\Bx)\Be_0, \eeq{1.11}
and express the fields in the form
\beq \Be^s(\Bx)=\Be_1(\Bx)+\Be_2(\Bx), \quad \Bd^s(\Bx)=\Bd_1(\Bx)+\Bd_2(\Bx)\quad {\rm with}~\Be_1,\Bd_1\in \CV,\quad \Be_2,\Bd_2\in \CH. \eeq{1.12}
The projections onto $\CV$ and $\CH$ are $\BGP_1$ and $\BGP_2$ whose action on a field $\Bp(\Bx)\in \CK$ are given by
\beq \BGP_1\Bp=\Gc\lang \Bp\rang, \quad \BGP_2\Bp=\Bp-\Gc\lang \Bp\rang. \eeq{1.13}
Applying $\BGP_2$ to both sides of \eq{1.12} gives
\beq \Bd_2=\BGP_2\BGe(\Bx)\Be_2=\BGe(\Bx)\Be_2, \eeq{1.14}
while applying $\BGP_1$ to both sides of \eq{1.12}, or equivalently subtracting \eq{1.14} from it, gives 
\beq \Bd_1=\BGe_1\Be_1+(\BGe_1-\BGe_0)\Gc(\Bx)\Be_0. \eeq{1.15}
The equations \eq{1.6}, \eq{1.12}, and \eq{1.14} are the defining equations for a $Y$-problem: given $\Be_1\in\CV$, find $\Bd_1\in \CV$ and 
$\Be_2,\Bd_2\in \CH$,
with $\Bd_2=\BGe\Be_2$ such that $\Be_1+\Be_2\in\CE$ and $\Bd_1+\Bd_2\in\CJ$. Since $\Bd_1$ depends linearly on $\Be_1$ we may write 
\beq \Bd_1=-\BY_*\Be_1, \eeq{1.16}
which defines the effective $Y$-tensor, $\BY_*$. Substituting this in \eq{1.15} gives
\beq \Be_1=-(\BY_*+\BGe_1)^{-1}(\BGe_1-\BGe_0)\Gc(\Bx)\Be_0. \eeq{1.17}
Also, by definition of the polarizability tensor $\BGa$,
\beq \BGa\Be_0=|\GO|\lang (\BGe_1-\BGe_0)(\Be_0+\Be^s(\Bx))\rang=|\GO||\lang(\BGe_1-\BGe_0)(\Be_0+\Be_1)\rang,
\eeq{1.18}
and so we see that
\beq \BGa=|\GO|[(\BGe_1-\BGe_0)-(\BGe_1-\BGe_0)(\BY_*+\BGe_1)^{-1}(\BGe_1-\BGe_0)]. \eeq{1.19}

\section{Bounds on the orientationally averaged complex polarizability tensor}
\setcounter{equation}{0}
Bounds on the polarizabilty tensor are an obvious consequence of bounds on the effective dielectric constant
of composite materials. Consider an inclusion $\GO$ of volume $|\GO|$ having isotropic dielectic constant $\Ge_1$ which is surrounded by
material with dielectric constant $\Ge_2=1$. We let $\Gc_1=\Ge_1-1$ denote the susceptability of phase 1 (that is not to be confused with the
characteristic function $\Gc(\Bx)$). Let $\BGa$ be its (possibly anisotropic) polarizability tensor. We then consider a 
dilute suspension of copies of this inclusion, with equally distributed random orientations.
Then insert material (or void) with dielectic constant $\Ge_2=1$ outside the inclusions.
By symmetry this material has an isotropic effective dielectric constant $\Ge_*$, which remains isotropic no matter what value the volume
fraction $p=|\GO|/\ell^3$ occupied by the inclusions happens to be. In the limit $p$ approaches zero one has the asymptotic formula 
\beq \Ge_*\approx 1+p\Tr(\BGa)/(d|\GO|), \eeq{2.1}
where $\Tr(\BGa)$ represents the average polarizability of the inclusions (in which $\Tr(\BGa)$ denotes the sum of the diagonal elements of the polarizability tensor),
and  $d=2$ or $3$ is the dimensionality of the space.

If $\Ge_1$ is real, then Hashin-Shtrikman \cite{Hashin:1962:VAT} established that
the effective dielectric constant $\Ge_*$ lies between the formulae
\beq 1+\frac{dp(\Ge_1-1)}
{d+(1-p)(\Ge_1-1)},~~~~\Ge_1+\frac{3(1-p)\Ge_1(1-\Ge_1)}
{d\Ge_1+p(1-\Ge_1)}, \eeq{2.2}
where $d=2$ or $3$ is the dimensionality of the composite. Taking the limit $p\to 0$ of each expression and using \eq{2.1} establishes that
$\Tr(\BGa)/(d|\GO|)$ must lie between the bounds
\beq \Gc_1-\Gc_1^2/(d(1+\Gc_1)),\quad  \Gc_1-\Gc_1^2/(\Gc_1+d). \eeq{2.3}

If $\Ge_1$ is complex, then the Bergman-Milton \cite{Milton:1979:TST,Milton:1980:BCD,Milton:1981:BCP,Bergman:1980:ESM,Bergman:1982:RBC}
bounds imply that $\Ge_*$ lies inside the region of the complex plane bounded by the circular arcs
inscribed by the points
\beqa
\Ge^{BM}_1(v) & = & 1+p(\Ge_1-1)
-\frac{p(1-p)(\Ge_1-1)^2}
{(1-p)\Ge_1 + p +(d-1)(v/\Ge_1+(1-v))^{-1}}, \nonum
\Ge^{BM}_2(w) & = & 1+p(\Ge_1-1)
-\frac{p(1-p)(\Ge_1-1)^2}{(1-p)\Ge_1+ p +(d-1)(w\Ge_1+(1-w))},
\eeqa{2.4}
as the real parameters $v$ and $w$ vary along the real axis
between 0 and 1. Taking the limit $p\to 0$ of each expression and using \eq{2.1} establishes that
$\Tr(\BGa)/(d|\GO|)$ must lie inside the region of the complex plane bounded by the circular arcs inscribed by the points
\beq \Ga^{BM}_1(v)=\Gc_1
-\frac{\Gc_1^2}
{1+\Gc_1 +(d-1)(v/(1+\Gc_1)+(1-v))^{-1}},\quad
\Ga^{BM}_2(w)=\Gc_1
-\frac{\Gc_1^2}{1+\Gc_1+(d-1)(w\Gc_1+1)},
\eeq{2.5}
as the real parameters $v$ and $w$ vary along the real axis
between 0 and 1. The bounds \eq{2.5} imply the bounds of  Miller, Hsu, Homer Reid, DeLacy, Joannopoulos, Solja{\v{c}}i{\'c} and Johnson \cite{Miller:2014:FLE} 
on the quasistatic absorption of small particles (Owen Miller, private communication). They point out the relevance of these bounds to determining limits on the absortion of
light by smoke screens of small metal particles. 

In two-dimensions improved bounds were obtained by Milton \cite{Milton:1980:BCD,Milton:1981:BCP} who found that $\Ge_*$ lies inside 
the region of the complex plane bounded by the circular arcs
inscribed by the points
\beq \Ge^{M}_1(v)=\frac{(p\Ge_1+1-p+\Ge_1)(\Ge_1+1)
-(1-p)v(\Ge_1-1)^2}{((1-p)\Ge_1+p+1)(\Ge_1+1)
-(1-p)v(\Ge_1-1)^2},\quad
\Ge^{M}_2(w)=\Ge_1\frac{(p\Ge_1+2-p)(\Ge_1+1)
-pw(\Ge_1-1)^2}{((1-p)\Ge_1+p+\Ge_1)(\Ge_1+1)
-pw(\Ge_1-1)^2}.
\eeq{2.6}
Taking the limit $p\to 0$ of each expression and using \eq{2.1} establishes that
$\Tr(\BGa)/(2|\GO|)$ must lie inside the region of the complex plane bounded by the circular arc and straight line inscribed by the points
\beq \Ga^{M}_1(v)=\frac{2\Gc_1(2+\Gc_1)}{(2+\Gc_1)^2-v\Gc_1^2},\quad
\Ga^{M}_2(w)=\frac{\Gc_1(2+\Gc_1)}{2(1+\Gc_1)}-\frac{w\Gc_1^3}{(\Gc_1+1)(\Gc_1+2)},
\eeq{2.7}
as the real parameters $v$ and $w$ vary along the real axis
between 0 and 1. 

This extremely simple approach to deriving bounds on the polarizabilty tensor is entirely rigorous once the asymptotic formula \eq{2.1} is established. By this method
rigorous bounds on the real and complex polarizabilty $\Ga$ of two-dimensional inclusions having an isotropic polarizability tensor were established in 
figures 2 and 3 of Milton, McPhedran, and McKenzie \cite{Milton:1981:TPA}, reproduced here in Figure 1,
and it was noted that when $\Ge_1$ is real the bounds are sharp for a disk, and for a very thin annulus. For two-dimensional inclusions that are perfectly 
conducting (effectively with $\Ge_1$ being infinite) P\'olya and Szeg\"o \cite{Polya:1951:IIM} had shown that the circular disk has the lowest average polarizabilty
of any inculsion shape of the same area, where the average is taken over all orientations, and they conjectured that a perfectly conducting sphere had the lowest average polarizabilty
of any inculsion shape of the same area. The conjecture is proved by the Hashin-Shtrikman bounds \eq{2.3}. A stronger form of the conjecture states that the sphere is the
only shape that attains the bounds: this and the related weak Eshelby conjecture were proved in \cite{Kang:2008:SPS} -- 
see also \cite{Liu:2008:SEC} for an independent proof of the weak Eshelby conjecture
that states an ellipsoid is the only shape inside which the field is uniform for all uniform applied fields. 

When $\Ge_1$ is real tighter bounds on the anisotropic polarizability tensor $\BGa$ (without averaging over orientations)
were obtained by Lipton \cite{Lipton:1993:IEE} by considering a dilute array of inclusions all with the same orientation, having an 
effective dielectric tensor $\BGe_*\approx \BI+p\BGa/|\GO|$ in the limit $p\to 0$. Lipton obtained the polarizabilty bounds by substituting this 
expression in the Tartar--Murat--Lurie-Cherkaev bounds \cite{Murat:1985:CVH,Tartar:1985:EFC,Lurie:1982:AEC,Lurie:1984:EEC}, and taking the limit $p\to 0$.

Lipton \cite{Lipton:1993:IEE} similarly derived bounds on the average elastic polarizability tensor, averaged over an ensemble of grain orientations not necessarily distributed randomly
with the inclusion and matrix having real moduli, from the low volume fraction limit of the bounds of Avellaneda \cite{Avellaneda:1987:OBM}
and noted they were sharp for suitable distributions of plate-like inclusions with at most 15 orientations in three dimensions (more recent work of 
\cite{Francfort:1995:FOM} implies that 6 orientations suffice). Shape independent bounds
on the average elastic polarizability tensor also follow by taking the low volume fraction limit of the ``Trace bounds'' of \cite{Milton:1988:VBE}
and Zhikov \cite{Zhikov:1988:ETA, Zhikov:1991:ETA}. Capdeboscq and Kang \cite{Capdeboscq:2008:IHS} show these can be tightened for inclusions which have some local
thickness. 
 
When the bulk modulus $\Gk_1$ and shear modulus $\Gm_1$ of the given inclusion phase are complex, while the bulk modulus 
$\Gk_0$ and shear modulus $\Gm_0$ of the surrounding material are real, then one can again consider the complex effective bulk modulus $\Gk_*$ and the complex
effective shear modulus $\Gm_*$ of a dilute suspension of copies of the inclusion, randomly orientated,  and occupying a volume fraction $p$ tending to zero. 
The available bounds on $\Gk_*$
and $\Gm_*$ are naturally expressed in terms of their $Y$-transforms,
\beqa y_{\Gk}& = & -(1-p)\Gk_1-p\Gk_0+\frac{p(1-p)(\Gk_1-\Gk_0)^2}
{p\Gk_1+ (1-p)\Gk_0-\Gk_*}, \nonum
 y_{\Gm}& = & -(1-p)\Gm_1-p\Gm_0+\frac{p(1-p)(\Gm_1-\Gm_0)^2}
{p\Gm_1+ (1-p)\Gm_0-\Gm_*}. \eeqa{2.8}
When the volume fraction $p$ is small we have
\beq \Gk_*\approx (1+p\Ga_\Gk/|\GO|)\Gk_0,\quad \Gm_*\approx (1+p\Ga_\Gm/|\GO|)\Gm_0, \eeq{2.9}
in which $\Ga_\Gk$ and $\Ga_\Gm$ are the average bulk and shear polarizabilities, where these are obtained by averaging the possibly anisotropic
fourth-order elastic polarizability tensor of the given inclusion over all orientations.
Substituting these expressions in \eq{2.8} we see that in the limit $p\to 0$ the formulae for $y_{\Gk}$ and $y_{\Gm}$ reduce to
\beq y_{\Gk}=-\Gk_1+\frac{(\Gk_1-\Gk_0)^2}{\Gk_1-\Gk_0(1+\Ga_\Gk/|\GO|)},\quad
y_{\Gm}=-\Gm_1+\frac{(\Gm_1-\Gm_0)^2}{\Gm_1-\Gm_0(1+\Ga_\Gm/|\GO|)}.
\eeq{2.10}
Then the Berryman--Gibiansky--Milton bounds on $y_{\Gk}$ and $y_{\Gm}$ for viscoelastic media \cite{Gibiansky:1993:EVM,Milton:1997:EVM,Gibiansky:1999:EVM}, that were derived using the Cherkaev--Gibiansky variational principles
\cite{Cherkaev:1994:VPC}, with $y_{\Gk}$ and $y_{\Gm}$ replaced by the expressions \eq{2.10},
directly give bounds on the possible complex values of the average bulk and shear polarizabilities
$\Ga_\Gk$ and $\Ga_\Gm$.
\section{Acoustic scattering}
\setcounter{equation}{0}
The polarizability problem is of course a limiting case of the scattering problem when the frequency of the incident field is very low, 
so that the wavelength of the incident radiation is much larger than the inclusion size. The success in
section 2 in reposing this as a $Y$-problem suggests that we might be able to repose acoustic scattering at any 
frequency as a $Y$-problem by eliminating the incident fields from the equations.

Let $P^a(\Bx)$ and $\Bv^a(\Bx)$ be the plain wave pressure and velocity fields that solve the acoustic equations in a homogeneous medium
with density $\Gr_0$ and bulk modulus $\Gk_0$, i.e.
\beq
\underbrace{\begin{pmatrix}-i\Bv^a \\ -i\Div\Bv^a \end{pmatrix}}_{\CG^a}=
\underbrace{\begin{pmatrix}-(\Go\Gr_0)^{-1}\BI_d & 0 \\ 0 & \Go/\Gk_0\end{pmatrix}}_{\BZ_0}
\underbrace{\begin{pmatrix}\Grad P^a \\ P^a \end{pmatrix}}_{\CF^a},
\eeq{3.1}
where $\BI_d$ is the $d\times d$ identity matrix.
Specifically, if $P^a(\Bx)=p^ae^{i\Bk_0\cdot\Bx}$ then these have the solution
\beq \CF^a=\begin{pmatrix}\Grad P^a \\ P^a \end{pmatrix}=\begin{pmatrix} i\Bk_0 p^a \\ p^a \end{pmatrix}e^{i\Bk_0\cdot\Bx}, \quad \quad
\CG^a=\begin{pmatrix}-i\Bv^a \\ -i\Div\Bv^a \end{pmatrix}=\begin{pmatrix} -ip^a\Bk_0/(\Go\Gr_0) \\ p^a\Go/\Gk_0 \end{pmatrix}e^{i\Bk_0\cdot\Bx},
\eeq{3.2}
implying that $v^a=p^a\Bk_0/(\Go\Gr_0)e^{i\Bk_0\cdot\Bx}$ and that $\Bk_0$ must have magnitude $k_0=|\Bk_0|$ given by
\beq k_0=\sqrt{\Go^2\Gr_0/\Gk_0}. \eeq{3.3}
We define $\CV^0$ as the space spanned by all fields of the form
\beq \Gc(\Bx)\begin{pmatrix} a_1\Bk_0 \\ a_2\end{pmatrix}e^{i\Bk_0\cdot\Bx}, \eeq{3.3a}
as the complex constants $a_1$ and $a_2$ vary and $\Bk_0$ varies, with $k_0=|\Bk_0|$ fixed and given by \eq{3.3}.
We emphasize that fields in $\CV^0$ do not necessarily have the form \eq{3.3a} but rather are a linear sum of fields of this form.
The space $\CV^0$ is the space of fields that exist inside the inclusion when it has the same properties as the matrix, and therefore
is the analog of the space $\CV$ in the polarizability problem. 

Given fields 
\beq \CP(\Bx)=\begin{pmatrix} \Bp(\Bx) \\ p(\Bx) \end{pmatrix},\quad \CP'(\Bx)=\begin{pmatrix} \Bp'(\Bx) \\ p'(\Bx) \end{pmatrix},\eeq{3.5a}
where $\Bp(\Bx)$ and $\Bp'(\Bx)$ are $d$-dimensional vector fields, and $p(\Bx)$ and $p'(\Bx)$ are scalar fields, 
we define the inner product
\beq (\CP',\CP)=\lim_{r_0\to\infty}\int_{t} w(t)(\CP',\CP)_{r_0t}\,\,dt, \eeq{3.5aa}
in which $w(t)$ is some smooth nonnegative weighting function, with say the properties that
\beq w(t)=0~~{\rm when}~~ t\leq 1/2~~{\rm or}~~t\geq 2,~~{\rm and}~~1=\int_{1/2}^2w(t)dt, \eeq{3.5ab}
and
\beq
(\CP',\CP)_r=\int_{B_r}\overline{\CP'(\Bx)}\cdot\CP(\Bx),\quad{\rm where}~
\overline{\CP'(\Bx)}\cdot\CP(\Bx)\equiv\overline{\Bp'(\Bx)}\cdot\Bp(\Bx)+\overline{p'(\Bx)}p(\Bx),\,\eeq{3.5b}
and $\overline{a}$ denotes the complex conjugate of $a$ for any quantity $a$. We define $\CK^0$ as the space of 
fields $\CP^0$ such that the norm $|h\CP^0|=(h\CP^0,h\CP^0)^{1/2}$, with inner product given by \eq{3.5aa}, is finite for all scalar functions $h(\Bx)\in C_0^\infty(\R^d)$
(where $C_0^\infty(\R^d)$ is the set of all infinitely differentiable functions with compact support) and which additionally have the asymptotic behavior
\beq \CP^0(\Bx)=\frac{e^{ik_0|\Bx|}}{|\Bx|}\left\{\begin{pmatrix} \widehat{\Bx}R^s_\infty({\widehat{\Bx}}) \\ S^s_\infty({\widehat{\Bx}}) \end{pmatrix}+\mathcal{O}\left(\frac{1}{|\Bx|}\right)\right\}
+\frac{e^{-ik_0|\Bx|}}{|\Bx|}\left\{\begin{pmatrix} \widehat{\Bx}R^i_\infty({\widehat{\Bx}}) \\ S^i_\infty({\widehat{\Bx}})  \end{pmatrix}+\mathcal{O}\left(\frac{1}{|\Bx|}\right)\right\},
\eeq{3.baa}
for some complex scalar functions $R^s_\infty(\Bn)$, $S^s_\infty({\widehat{\Bx}})$, $R^i_\infty(\Bn)$  and $S^i_\infty(\Bn)$ defined on the unit sphere $|\Bn|=1$,  where $\widehat{\Bx}=\Bx/|\Bx|$. Here
the superscript $s$ is used because these field components will later be associated with the scattered field. The superscript $i$ is used because these field components will later be associated with 
incoming fields, though not the incoming fields associated with the incident fields $P^a$ and $\Bv^a$ as these will be treated separately. The subspace $\CK^0$ has been defined in this way
to ensure that if $\CP\in\CK^0$ then so are its real and imaginary parts, as taking real and imaginary parts are crucial to developing variational principles along the lines
first suggested by Cherkaev and Gibiansky \cite{Cherkaev:1994:VPC}. 
We define $\CK^s$ as the space of fields $\CP^0\in\CK^0$ satisfying the 
condition that $R^i_\infty(\Bn)=S^i_\infty(\Bn)=0$ for all $\Bn$. Note that the norm $|\CP^0|=(\CP^0,\CP^0)^{1/2}$ is not finite for fields in $\CK^0$ 
if $R^s_\infty(\Bn)$, $S^s_\infty({\widehat{\Bx}})$, $R^i_\infty(\Bn)$  or $S^i_\infty(\Bn)$ is nonzero. We define $\CH^0$ as the orthogonal complement of $\CV^0$ in the space $\CK^0$.

We are interested in solving
\beq
\underbrace{\begin{pmatrix}-i(\Bv^a+\Bv^s) \\ -i\Div(\Bv^a+\Bv^s) \end{pmatrix}}_{\CG^a+\CG^s}=
\underbrace{\begin{pmatrix}-(\Go\BGr)^{-1} & 0 \\ 0 & \Go/\Gk\end{pmatrix}}_{\BZ(\Bx)}
\underbrace{\begin{pmatrix}\Grad (P^a+P^s) \\ (P^a+P^s) \end{pmatrix}}_{\CF^a+\CF^s},
\eeq{3.4}
where $P^s(\Bx)$ is the scattered pressure, $\Bv^s(\Bx)$ the associated scattered velocity, and $\CG^s,\CF^s\in\CK^s$. Here the density
matrix $\BGr$ takes the value $\Gr_0\BI_d$ outside the inclusion and the value $\BGr_1$ inside the inclusion, while the bulk
modulus scalar $\Gk$ takes the value $\Gk_0$ outside the inclusion and the value $\Gk_1$ inside the inclusion. Due to viscoelasticity
(energy loss under oscillatory compression) it is quite natural to have a bulk modulus that is complex with a negative imaginary part.
We also allow for the density $\BGr_1$ to depend on the frequency $\Go$ and be anisotropic and possibly complex valued with a positive 
imaginary part, even with a negative real part, since this can be the case in metamaterials 
\cite{Schoenberg:1983:PPS,Auriault:1985:DCE,Auriault:1994:AHM,Zhikov:2000:EAT,Sheng:2003:LRS,Movchan:2004:SRR,Liu:2005:AMP,Milton:2006:CEM,Milton:2007:MNS,Torrent:2008:AMD, Smyshlyaev:2009:PLE,Buckmann:2015:EMU}. Alternatively, one can consider 
electromagnetic scattering off a cylindrical shaped inclusion (not necessarily with a circular cross-section) and then the transverse
electric and transverse magnetic equations are directly analogous to the two-dimensional acoustic equations. In that context it is
well known that both the electric permittivity tensor and magnetic permeability tensor can be anisotropic and complex valued, with positive
semidefinite imaginary parts.
 
Now using the relation \eq{3.1}, that $\CG^a=\BZ_0\CF^a$, we rewrite \eq{3.4} as
\beq \CG^s(\Bx)=\BZ(\Bx)\CF^s(\Bx)+(\BZ_1-\BZ_0)\Gc(\Bx)\CF^a, \eeq{3.5}
in which $\Gc(\Bx)\CF^a\in\CV^0$. We define $\CE^0$ as the space of all fields $\CF^0$
in $\CK^0$ of the form
\beq \CF^0=\begin{pmatrix}\Grad P^0(\Bx) \\ P^0(\Bx) \end{pmatrix}, \eeq{3.5ba}
for some scalar field $P^0(\Bx)$, and we define $\CJ^0$  as the space of all fields $\CG^0$
in $\CK^0$ of the form
\beq\CG^0= \begin{pmatrix}-i\Bv^0 \\ -i\Div\Bv^0 \end{pmatrix}, \eeq{3.5bb}
for some vector field $\Bv^0(\Bx)$. The fields $\CF^0$ and $\CG^0$, being in $\CK^0$, have the asymptotic 
forms
\beqa \CF^0(\Bx)& = &\frac{e^{ik_0|\Bx|}}{|\Bx|}\left\{P^s_\infty({\widehat{\Bx}})\begin{pmatrix} ik_0\widehat{\Bx} \\ 1 \end{pmatrix}+\mathcal{O}\left(\frac{1}{|\Bx|}\right)\right\}
+\frac{e^{-ik_0|\Bx|}}{|\Bx|}\left\{P^i_\infty({\widehat{\Bx}})\begin{pmatrix} -ik_0\widehat{\Bx} \\ 1 \end{pmatrix}+\mathcal{O}\left(\frac{1}{|\Bx|}\right)\right\},\nonum
\CG^0(\Bx) & = & \frac{e^{ik_0|\Bx|}}{|\Bx|}\left\{V^s_\infty({\widehat{\Bx}})\begin{pmatrix} -i\widehat{\Bx}/k_0 \\ 1 \end{pmatrix}+\mathcal{O}\left(\frac{1}{|\Bx|}\right)\right\}
+\frac{e^{-ik_0|\Bx|}}{|\Bx|}\left\{V^i_\infty({\widehat{\Bx}})\begin{pmatrix} i\widehat{\Bx}/k_0 \\ 1 \end{pmatrix}+\mathcal{O}\left(\frac{1}{|\Bx|}\right)\right\},
\eeqa{3.5bc}
implying, through \eq{3.5ba} and \eq{3.5bb}, that at large $|\Bx|$,
\beqa P^0(\Bx)& \approx & \frac{e^{ik_0|\Bx|}}{|\Bx|}P^s_\infty(\widehat{\Bx})+\frac{e^{-ik_0|\Bx|}}{|\Bx|}P^i_\infty(\widehat{\Bx}), \nonum
\Bv^0(\Bx)&\approx&\widehat{\Bx}\frac{e^{ik_0|\Bx|}}{k_0|\Bx|}V^s_\infty(\widehat{\Bx})-\widehat{\Bx}\frac{e^{-ik_0|\Bx|}}{k_0|\Bx|}V^i_\infty(\widehat{\Bx}),
\eeqa{3.5c}
The Sommerfeld radiation condition in fact implies that the $P^i_\infty(\widehat{\Bx})$ and $V^i_\infty(\widehat{\Bx})$ associated with
the actual scattered pressure $P^s(\Bx)$ and scattered velocity, $\Bv^s(\Bx)$ are zero, but we keep these terms 
as we want to impose a ``constitutive law at infinity'' that forces $P^i_\infty(\widehat{\Bx})$ and $V^i_\infty(\widehat{\Bx})$ to be zero and thus
replaces the Sommerfeld radiation condition. Also we want to define the spaces $\CE^0$ and $\CJ^0$ so that if
$\CF^0\in\CE^0$ and $\CG^0\in\CJ^0$ then so are their real and imaginary parts. 
We extend the definition of $P^s_\infty(\widehat{\Bx})$, and $V^s_\infty(\widehat{\Bx})$
to all of $\R^3$ except the origin in the natural way by letting
\beq P^s_\infty(\Bx)=P^s_\infty(\Bx/|\Bx|),\quad V^s_\infty(\Bx)=V^s_\infty(\Bx/|\Bx|). \eeq{3.5d}
Then using the fact that $|\Bx|=\sqrt{\Bx\cdot\Bx}$ and $\widehat{\Bx}=\Bx/\sqrt{\Bx\cdot\Bx}$ we obtain
\beqa \Grad P^s(\Bx) & = & \frac{ik_0\Bx e^{ik_0|\Bx|}}{|\Bx|^2}P^s_\infty(\widehat{\Bx})
-\frac{\Bx e^{ik_0|\Bx|}}{|\Bx|^{3}}P^s_\infty(\widehat{\Bx})
+\frac{e^{ik_0|\Bx|}}{|\Bx|^2}\Bp^s(\Bx/|\Bx|), \nonum
\Div\Bv^s(\Bx) & = & \frac{ie^{ik_0|\Bx|}}{|\Bx|}V^s_\infty(\widehat{\Bx})
+(d-2)\frac{e^{ik_0|\Bx|}}{k_0|\Bx|^2}V^s_\infty(\widehat{\Bx})
+\frac{e^{ik_0|\Bx|}}{k_0|\Bx|^2}v^s(\Bx/|\Bx|),
\eeqa{3.5e}
where
\beq \Bp^s(\Bx/|\Bx|)=|\Bx|\Grad P^s_\infty(\Bx),\quad v^s(\Bx/|\Bx|)=\Bx\cdot\Grad V^s_\infty(\Bx)  \eeq{3.5f}
only depend on $\Bx/|\Bx|$, since $\Grad P^s_\infty(\Gl\Bx)=(1/\Gl)\Grad P^s_\infty(\Gl\Bx)$ 
and $\Grad V^s_\infty(\Gl\Bx)=(1/\Gl)\Grad V^s_\infty(\Gl\Bx)$ for all real $\Gl>0$. 
The dominant terms in the expressions in \eq{3.5e} are the first terms, which justifies those terms in \eq{3.5bc}
that involve $P^s_\infty({\widehat{\Bx}})$ and $V^s_\infty({\widehat{\Bx}})$. The terms that involve $P^i_\infty({\widehat{\Bx}})$ and $V^i_\infty({\widehat{\Bx}})$
are justified in a similar way by extending those functions to all of $\R^3$ except the origin.
Using integration by parts we have the key identity that
\beq (\CF^0,\CG^0)_r=\int_{B_{r}}\overline{\CF^0}\cdot\CG^0\,d\Bx = \int_{\Md B_{r}}-i\overline{P^0(\Bx)}\Bn\cdot\Bv^0(\Bx)\,dS.
\eeq{3.5faa}
From \eq{3.5c} we see that when $|\Bx|$ is large,
\beq \overline{P^0(\Bx)}\Bn\cdot\Bv^0(\Bx)\approx 
\frac{1}{k_0|\Bx|^2}\left[\overline{P^s_\infty({\widehat{\Bx}})}V^s_\infty({\widehat{\Bx}})-\overline{P^i_\infty({\widehat{\Bx}})}V^i_\infty({\widehat{\Bx}})
+e^{2ik_0r}\overline{P^i_\infty({\widehat{\Bx}})}V^s_\infty({\widehat{\Bx}})-e^{-2ik_0r}\overline{P^s_\infty({\widehat{\Bx}})}V^i_\infty({\widehat{\Bx}})\right].
\eeq{3.5fab}
The last two cross terms that involve $e^{2ik_0r}$ and $e^{-2ik_0r}$ obviously oscillate very rapidly with $r$ and will average to zero
in the integral \eq{3.5aa} involving the smooth weight function $w(t)$. Thus we get
\beqa (\CF^0,\CG^0)& = & \lim_{r_0\to\infty}\int_{t}dt \frac{-iw(t)}{k_0r_0^2t^2}\int_{B_{r_0t}}
\left[\overline{P^s_\infty({\widehat{\Bx}})}V^s_\infty({\widehat{\Bx}})-\overline{P^i_\infty({\widehat{\Bx}})}V^i_\infty({\widehat{\Bx}})\right]\,dS \nonum
& = &\frac{-i}{k_0}\int_{B_{1}}
\left[\overline{P^s_\infty(\Bn)}V^s_\infty(\Bn)-\overline{P^i_\infty(\Bn)}V^i_\infty(\Bn)\right]\,dS,
\eeqa{3.5fa}
This lack of orthogonality of the subspaces $\CE^0$ and $\CJ^0$ can be remedied by introducing an auxillary space $\CA$ of 
two-component vector fields $\Bq(\Bn)=[q_1(\Bn),q_2(\Bn)]$ defined, and square integrable, on the unit sphere $|\Bn|=1$. 
We then consider the Hilbert space $\CK$ composed of fields $[\CP,q_1,q_2]$, where $\CP\in\CK^0$ and $\Bq(\Bn)=[q_1(\Bn),q_2(\Bn)]\in\CA$.
In general, the field components $q_1(\Bn)$ and $q_2(\Bn)$ need not be related to the functions $R^s_\infty(\Bn)$, $S^s_\infty({\widehat{\Bx}})$, $R^i_\infty(\Bn)$  and $S^i_\infty(\Bn)$
appearing in the asymptotic expansion \eq{3.baa}. The inner product between two fields $\CQ=[\CP,q_1,q_2]$ and $\CQ'=[\CP',q_1',q_2']$ in $\CK$ is defined as
\beq \lang\CQ',\CQ\rang =(\CP',\CP)+\frac{1}{2k_0}\int_{|\Bn|=1}\overline{q'_1(\Bn)}q_1(\Bn)+\overline{q'_2(\Bn)}q_2(\Bn)\,dS. \eeq{3.5fb}
We define $\CE$ to consist of fields $\CF=[\CF^0,-iP^s_\infty+iP^i_\infty, P^s_\infty+P^i_\infty]\in\CK$, where $\CF^0\in\CE^0$ while $P^s_\infty(\Bn)$ 
and $P^i_\infty(\Bn)$ are those functions that enters its asymptotic form \eq{3.5bc}. 
We define $\CJ$ to consist of fields $\CG=[\CG^0, V^s_\infty+V^i_\infty, iV^s_\infty-iV^i_\infty]\in\CK$, where $\CG^0\in\CJ^0$ while $V^s_\infty(\Bn)$ 
and $V^i_\infty(\Bn)$ are those functions that enters its asymptotic form \eq{3.5bc}. In each case the accompanying auxillary fields are respectively
\beq \Bq_{\CF}(\Bn)=[-iP^s_\infty(\Bn)+iP^i_\infty(\Bn), P^s_\infty(\Bn)+P^i_\infty(\Bn)],~~{\rm and}~~ 
\Bq_{\CG}(\Bn)=[V^s_\infty(\Bn)+V^i_\infty(\Bn), iV^s_\infty(\Bn)-iV^i_\infty(\Bn)].
\eeq{3.5fba}
The auxillary fields have been defined in this way, in part, to ensure that if $\CF\in\CJ$ and $\CG\in\CJ$ then so too do their real and imaginary
parts lie in these subspaces. 

Now the inner product of $\CF$ and $\CG$ is 
\beq \lang \CF,\CG\rang=(\CF^0,\CG^0)+\frac{1}{k_0}\int_{\Md B_{1}}i\overline{P^s_\infty(\Bn)}V^s_\infty(\Bn)\,dS
+\frac{1}{k_0}\int_{\Md B_{1}}-i\overline{P^i_\infty(\Bn)}V^i_\infty(\Bn)\,dS
=0, 
\eeq{3.5fc}
which implies the orthogonality of the spaces $\CE$ and $\CJ$. Similarly we extend the definition of $\CV^0$: $\CV$ consists of 
pairs $\CP_1=[\CP_1^0,0,0]$ where $\CP_1^0\in\CV^0$. We define $\CH^0$ as the orthogonal complement of $\CV^0$ in the space $\CK^0$, 
and $\CH$ as the orthogonal complement of $\CV$ in the space $\CK$:
it consists of fields $\CP_2=[\CP_2^0,q_1,q_2]$, where $\Bq=[q_1,q_2]\in\CA$ and $\CP_2^0\in\CH^0$, which implies
\beq \int_{\GO}\overline{\CP_2^0(\Bx)}\cdot\begin{pmatrix} a_1\Bk_0 \\ a_2\end{pmatrix}e^{i\Bk_0\cdot\Bx}\,d\Bx=0~~~{\rm for~all~}a_1,a_2.
\eeq{3.5fca}

The fields $\CF^s$ and $\CG^s$ that solve \eq{3.4} are respectively in $\CE^0$ and $\CJ^0$:
\beq \CF^s\in\CE^0,\quad\quad\CG^s\in\CJ^0,
\eeq{3.5g}
and we express them in the form
\beq \CF^s(\Bx)=\CF_1^s(\Bx)+\CF_2^s(\Bx), \quad \CG^s(\Bx)=\CG_1^s(\Bx)+\CG_2^s(\Bx),\quad {\rm with}~\CF_1^s,\CG_1^s\in \CV^0,\quad \CF_2^s,\CG_2^s\in \CH^0. \eeq{3.6}
Clearly we have
\beq \CG_1^s=\CG^a+\CG_1^s-\BZ_0\CF^a=\BZ_1(\CF^a+\CF_1^s)-\BZ_0\CF^a,
\eeq{3.7}
and substracting this formula for $\CG_1^s(\Bx)$,
\beq \CG_1^s(\Bx)=\BZ(\Bx)\CF_1^s(\Bx)+(\BZ_1-\BZ_0)\Gc(\Bx)\CF^a, \eeq{3.8}
from \eq{3.5} we see that
\beq \CG_2^s(\Bx)=\BZ(\Bx)\CF_2^s(\Bx). \eeq{3.9}
Of course since $\CF^s$ and $\CG^s$ lie in $\CK^s$, rather than just $\CK^0$, the asymptotic components $P^i_\infty(\Bn)$ and $V^i_\infty(\Bn)$ are zero.
However let us remove this restriction and allow nonzero values of $P^i_\infty(\Bn)$ and $V^i_\infty(\Bn)$, that we will then show must be zero.
The associated fields $\CF_2=[\CF^s_2,-iP^s_\infty+iP^i_\infty, P^s_\infty+P^i_\infty]\in\CH$ and 
$\CG_2=[\CG^s_2, V^s_\infty+V^i_\infty, iV^s_\infty-iV^i_\infty]\in\CH$ have auxillary components $\Bq_{\CF}$ and $\Bq_{\CG}$ given by \eq{3.5fba}.
We require that these auxillary components satisfy the constitutive law
\beq \Bq_{\CG}=\frac{i\Go}{\kappa_0}\Bq_{\CF}, \eeq{3.9a}
or equivalently, we have
\beq V^s_\infty+V^i_\infty=\Go(P^s_\infty-P^i_\infty)/\kappa_0,\quad iV^s_\infty-iV^i_\infty=i\Go(P^s_\infty+P^i_\infty)/\kappa_0.
\eeq{3.9ab}
Additionally, the constitutive law \eq{3.9} allows us to relate the asymptotic terms of $\CG_2^s(\Bx)$ and $\CF_2^s(\Bx)$ giving
\beq V^s_\infty=\Go P^s_\infty/\kappa_0,\quad V^i_\infty=\Go P^i_\infty/\kappa_0. \eeq{3.9ac}
In conjunction with \eq{3.9ab} this forces
\beq V^i_\infty(\Bn)= P^i_\infty(\Bn)=0, \eeq{3.9ad}
as desired. Thus we have replaced the Sommerfeld radiation condition with the constitutive law \eq{3.9a}.

There is a natural division of the Hilbert space $\CH$ into three orthonormal subpaces: the space $\CS_1$ of fields $\CP_2=[\CP^s_2,0,0]$ where
$\CP_2^s(\Bx)\in\CH^s$ is nonzero only in the inclusion phase; the space $\CS_2$ of fields $\CP_2=[\CP^s_2,0,0]$ where
$\CP_2^s(\Bx)\in\CH^s$ is nonzero only in the matrix phase; and the space $\CS_3$ of fields $\CP_2=[0,q_1,q_2]$ where $\Bq=[q_1,q_2]\in\CA$. In the first two cases
we define the action of an operator $\BL$ on these fields to be $\BL\CP_2=[\BZ_1\CP_2^s,0,0]$ and  $\BL\CP_2=[\BZ_0\CP_2^s,0,0]$ respectively,
and in the third case to be $\BL\CP_2=[0,(i\Go/\kappa_0)q_1,(i\Go/\kappa_0)q_2 ]$ to agree with \eq{3.9a}.
More generally, the action of $\BL$ on any field $\CP_2$ in $\CH$ is obtained
by resolving $\CP_2$ into its components in these three subspaces, and summing the action of $\BL$ on the component fields. With
this definition \eq{3.9} and \eq{3.9a} imply the constitutive law
\beq \CG_2=\BL\CF_2. \eeq{3.9b}

The equations \eq{3.5g}, \eq{3.6}, and \eq{3.9b} are the defining equations for a $Y$-problem: given $\CF_1\in\CV$, find $\CG_1\in \CV$ and 
$\CF_2,\CG_2\in \CH$,
with $\CG_2=\BL\CF_2$ such that $\CF_1+\CF_2\in\CE$ and $\CG_1+\CG_2\in\CJ$. Since $\CG_1$ depends linearly on $\CF_1$ we may write 
\beq \CG_1=-\BY_*\CF_1, \eeq{3.10}
which defines the effective $Y$-operator $\BY_*$. Since $\CV$ consists of fields $\CP_1=[\CP_1^s,0,0]$ where $\CP_1^s\in\CV^0$ we can equivalently
write the relation \eq{3.10} as
\beq \CG_1^s=-\BY_*^0\CF_1^s, \eeq{3.10a}
which defines the effective $Y$-operator $\BY_*^0$. Substituting this in \eq{3.8} gives
\beq \CF_1^s=-(\BY_*^0+\BZ_1)^{-1}(\BZ_1-\BZ_0)\Gc(\Bx)\CF^a. \eeq{3.11}
We emphasize that $\BY_*^0$ is a linear operator which maps fields in $\CV^0$ to fields in $\CV^0$: it is not a matrix that acts locally on the fields.
Notice that the orthogonality of the spaces $\CE$ and $\CJ$ and the orthogonality of the spaces $\CV$ and $\CH$ imply
\beq 0=\lang \CF_1+\CF_2,\CG_1+\CG_2\rang= \lang \CF_1,\CG_1\rang+\lang \CF_2,\CG_2\rang=-(\CF_1^s,\BY_*^0\CF_1^s)+\lang \CF_2,\BL\CF_2\rang.
\eeq{3.11a}
Now let 
\beq {P^a}'(\Bx)=e^{-i\Bk'_0\cdot\Bx} \quad {\rm and}~~ {\Bv^a}'(\Bx)=-i(\Go\Gr_0)^{-1}\Grad e^{-i\Bk'_0\cdot\Bx}=-\Bk'_0(\Go\Gr)^{-1} e^{-i\Bk'_0\cdot\Bx} \eeq{4.17}
be another plain wave pressure and associated velocity field that solve the acoustic equations in the homogeneous medium
with density $\Gr_0$ and bulk modulus $\Gk_0$, i.e.
\beq
\underbrace{\begin{pmatrix}-i{\Bv^a}' \\ -i\Div{\Bv^a}' \end{pmatrix}}_{{\CG^a}'}=
\underbrace{\begin{pmatrix}-(\Go\Gr_0)^{-1}\BI_d & 0 \\ 0 & \Go/\Gk_0\end{pmatrix}}_{\BZ_0}
\underbrace{\begin{pmatrix}\Grad {P^a}' \\ {P^a}' \end{pmatrix}}_{{\CF^a}'}.
\eeq{3.12}
Using the key identity we have that 
\beqa I_1 & \equiv & \int_{B_{r_0}}{\CF^a}'\cdot(\CG^s-\BZ_0\CF^s)\,d\Bx=\int_{B_{r_0}}({\CF^a}'\cdot\CG^s-(\BZ_0{\CF^a}') \cdot\CF^s)\,d\Bx \nonum
& = & \int_{B_{r_0}}({\CF^a}'\cdot\CG^s-{\CG^a}'\cdot\CF^s)\,d\Bx
=\int_{\Md B_{r_0}}-i{P^a}'\Bn\cdot\Bv^s+iP^s\Bn\cdot{\Bv^a}'\,dS.
\eeqa{3.13}
Clearly the integrand on the left hand side vanishes outside $\GO$ and so the integral must be independent of the radius $r$ of the ball $B_{r_0}$
(so long as it contains the inclusion). So one can evaluate this integral by taking the limit $r_0\to\infty$, which will be done in the next section. The
identity \eq{3.13} is the analog of the identity \eq{1.9} that for the polarization problem expresses an integral over the inclusion in terms of the
far-field. 

Alternatively, using \eq{3.8}, we can write the left hand side of the equation as
 \beqa I_1 & = & \int_{B_{r_0}}{\CF^a}'\cdot(\CG^s-\BZ_0\CF^s)\,d\Bx  =  \int_{\R^3}\Gc{\CF^a}'\cdot(\CG_1^s+\CG_2^s)-\Gc{\CF^a}'\cdot(\BZ_0\CF_1^s)-(\BZ_0\Gc{\CF^a}')\cdot\CF_2^s\,d\Bx \nonum
& = & \int_{\GO}{\CF^a}'\cdot(\BZ_1-\BZ_0)\CF_1+{\CF^a}'\cdot(\BZ_1-\BZ_0)\CF^a   \,d\Bx\nonum
& = & \int_{\GO}{\CF^a}'\cdot[(\BZ_1-\BZ_0)-(\BZ_1-\BZ_0)(\BY_*+\BZ_1)^{-1}(\BZ_1-\BZ_0)]\CF^a\,d\Bx,
\eeqa{3.14} 
where we have used the fact that $\Gc{\CF^a}'$ and $\BZ_0\Gc{\CF^a}'$ are in $\CV^0$, and hence orthogonal to $\CG_2^s$ and $\CF_2^s$. Let $\widetilde\CV$ be that
subspace of $\CV$ comprised of all linear combinations of fields of the form 
\beq \Gc(\Bx)\CF^a=\Gc(\Bx)\begin{pmatrix} ip^a\Bk_0 \\ p^a \end{pmatrix}e^{i\Bk_0\cdot\Bx}, \eeq{3.15}
as $p^a$ and $\Bk_0$ vary, with $k_0=|\Bk_0|$ fixed and given by \eq{3.3}. Clearly both $\Gc(\Bx)\CF^a$ and  $\Gc(\Bx){\CF^a}'$ lie in $\widetilde\CV$
so if $\widetilde\GP$ denotes the projection operator onto $\widetilde\CV$, we have the identity
\beq I_1=\lim_{r_0\to\infty}\int_{\Md B_{r_0}}-i{P^a}'\Bn\cdot\Bv^s+iP^s\Bn\cdot{\Bv^a}'\,dS=\int_{\GO}{\CF^a}'\cdot\BGL\CF^a\,d\Bx,
\eeq{3.16}
where $\BGL$ is the scattering operator
\beq
\BGL=\widetilde\GP[(\BZ_1-\BZ_0)-(\BZ_1-\BZ_0)(\BY_*^0+\BZ_1)^{-1}(\BZ_1-\BZ_0)]\widetilde\GP,
\eeq{3.17}
with this expression being analogous to the expression \eq{1.19} for the polarizability tensor. Thus the bilinear form
$I_1({\CF^a}',\CF^a)$ defines $\BGL$ and we will see in the next section that $I_1$ can be determined from the far field
scattering amplitudes $P^s_\infty(\Bn)$. 

\section{Expressing the scattered field in terms of integrals over the inclusion}
\labsect{farfield}
\setcounter{equation}{0}
Here our goal is to evaluate the integral on the right hand side of \eq{3.13} using the asymptotic formula,
\beq P^s(\Bx)=\frac{e^{ik_0|\Bx|}}{|\Bx|}P^s_\infty(\widehat{\Bx}),~~{\rm with}~ \widehat{\Bx}=\Bx/|\Bx|, \eeq{4.7}
for the scattered pressure field, and the associated asymptotic formula for the scattered velocity field 
$\Bv^s=-i(\Go\Gr_0)^{-1}\Grad P^s(\Bx)$. The calculation is the analog of the calculation \eq{1.10}, that expresses
a far field integral in terms of the dipole moment.

Suppose we take a ball $ B$ of radius $r$. Then the outwards unit 
normal to the ball surface is $\Bn=\Bx/r$ and consequently $\Bn\cdot\Bx=r$.
Using the fact that $|\Bx|=\sqrt{\Bx\cdot\Bx}$ and $\widehat{\Bx}=\Bx/\sqrt{\Bx\cdot\Bx}$ this gives
\beq \Bn\cdot\Grad P^s(\Bx)=\frac{\Md P^s(\Bx)}{\Md r}
\approx \frac{ik_0e^{ik_0r}}{r}P^s_\infty(\widehat{\Bx})
-\frac{e^{ik_0r}}{r^2}P^s_\infty(\widehat{\Bx}).
\eeq{4.15}
Hence at large distances, keeping $\widehat{\Bx}$ fixed the dominant term in the above expression for $\Bn\cdot\Grad P^s(\Bx)$ is the first term.
So just keeping this, we obtain
\beq  \Bn\cdot\Bv^s=-i(\Go\Gr_0)^{-1}\Bn\cdot\Grad P^s(\Bx)\approx (\Go\Gr_0)^{-1}\frac{k_0e^{ik_0r}}{r}P^s_\infty(\widehat{\Bx}).
\eeq{4.16}
Recall the pressure field ${P^a}'(\Bx)$ and associated velocity field ${\Bv^a}'(\Bx)$ are given by \eq{4.17}.
So, we need to evaluate
\beq I_1= \int_{\Md B_{r}}-i{P^a}'\Bn\cdot\Bv^s+iP^s\Bn\cdot{\Bv^a}'\,dS
\approx -i(\Go\Gr_0)^{-1}\int_{\Md B_{r}}e^{\Bk'_0\cdot\Bx}e^{ik_0r}P^s_\infty(\widehat{\Bx})(k_0+\Bn\cdot\Bk'_0)/r\,dS.
\eeq{4.17a}
Without loss of generality let us suppose that the $x_1$ axis has been chosen in the direction of $\Bk'_0$, so $e^{-i\Bk'_0\cdot\Bx}=e^{-ik_0x_1}$
and $\Bn\cdot\Bk'_0=k_0n_1=k_0x_1/r$. Let us use cylindrical coordinates $(x_1,\Gvr,\Gt)$ where $\Gvr=\sqrt{x_2^2+x_3^2}$ and $\tan\Gt=x_3/x_2$,
so that $x_2=\Gvr\cos\Gt$ and  $x_3=\Gvr\sin\Gt$. We then introduce the ratio $t=x_1/r$ and express
\beq P^s_\infty(\widehat{\Bx})=P^s_\infty(\Gt,t). \eeq{4.18}
Thus in cylindrical coordinates the far-field expression for the scattered pressure field at $\Md B$ becomes
\beq P^s(\Bx)\approx\frac{e^{ik_0r}}{r}P^s_\infty(\Gt,x_1/r). \eeq{4.18a}
We choose as our variables of integration the parameters $t=x_1/r$ and $\Gt$. In terms of $t$ and $\Gt$, we have
\beqa &~& x_1=rt,\quad \Gvr=r\sqrt{1-t^2},\quad e^{-i\Bk'_0\cdot\Bx}=e^{-ik_0rt},\quad (k_0+\Bn\cdot\Bk'_0)/r=k_0(1+t)/r,
\nonum
&~& dS=\Gvr\,d\Gt\,dx_1/\sqrt{n_2^2+n_3^2}=r^2\,d\Gt\,dt,
\eeqa{4.19}
where $n_2$ and $n_3$ are the components of the vector $\Bn=\Bx/r$. The only term in the integration that involves $\Gt$ is $P^s_\infty(\Gt,h)$,
so integrating this over $\Gt$ defines
\beq p_{\infty}(t)\equiv\int_{0}^{2\pi}P^s_\infty(\Gt,t)\,d\Gt. \eeq{4.20}
We obtain
\beq I_1\approx -i(\Go\Gr_0)^{-1}I_2,\quad I_2=\int_{-1}^1rf(t)e^{irg(t)}\,dt, \eeq{4.21}
where
\beq g(t) = -k_0t+k_0,\quad f(t) = k_0(1+t)p_{\infty}(t).
\eeq{4.22}
Asymptotic expressions in the limit $r\to\infty$ for integrals taking the form of $I_2$ in \eq{4.21} are available when $g(t)$ has a non-zero derivative
$g'(t)=-k_0$ for $1\geq t \geq -1$, which is clearly the case, 
and one has \cite{Iserles:2005:EQH},
\beq \lim_{r\to\infty}I_2=\frac{e^{irg(1)}f(1)}{ig'(1)}-\frac{e^{irg(-1)}f(-1)}{ig'(-1)}. \eeq{4.24}
We have $g'(1)=g'(-1)=-k_0$, while \eq{4.22} and \eq{4.18a} imply
\beq g(1)=0,\quad g(-1)= 2k_0, \quad f(1)=2k_0p_{\infty}(1)=4k_0\pi P^s_\infty(\Bk'_0/k_0), \quad f(-1)=0, \eeq{4.25}
that when substituted in \eq{4.24} gives
\beq  \lim_{r\to\infty}I_2=4i\pi P^s_\infty(\Bk'_0/k_0), \eeq{4.26}
which is independent of $\Ga$ as expected.  Hence we obtain an exact expression for $I_1$:
\beq I_1=4k_0\pi P^s_\infty(\Bk'_0/k_0)/(\Go\Gr_0). \eeq{4.27}

\section{A Minimization Variational Principle  for Acoustic Scattering}
\labsect{varprin}
\setcounter{equation}{0}
The fact that acoustic scattering can be regarded as a ``$Y$-problem'' naturally leads to
minimization variational principles. Here we follow the more or less standard approach
for deriving these variational principles (see \cite{Milton:1991:FER} and Section 19.6 of \cite{Milton:2002:TOC}),
using the transformation techniques of Gibiansky and Cherkaev \cite{Cherkaev:1994:VPC}. Some adaptation
is needed to allow for the fact that the matrix phase is lossless. This requires one to choose trial fields that solve the wave equation
exactly in the matrix phase.

For $\Bx$ in the inclusion phase we can take real and imaginary parts of the constitutive law \eq{3.9} to give
\beq \begin{pmatrix} \Real (\CG^s_2(\Bx)) \\ \Imag (\CG^s_2(\Bx))\end{pmatrix}
=\begin{pmatrix} -\BZ_1'' & \BZ_1' \\ \BZ_1' & \BZ_1'' \end{pmatrix}
\begin{pmatrix} \Imag (\CF^s_2(\Bx)) \\ \Real (\CF^s_2(\Bx))\end{pmatrix},
\eeq{5.1}
where $\BZ_1'$ and $\BZ_2''$ denote the real and imaginary parts of $\BZ_1=\BZ_1'+i\BZ_1''$. 
Let us begin by supposing that $\Go$ is real and that $1/\Gk_1$ and $\Gr$ have strictly positive imaginary parts
so that $\BZ_1''$ is a positive definite matrix. Then, following the ideas of Cherkaev and Gibiansky \cite{Cherkaev:1994:VPC}, that 
were extended to wave equations by Milton, Seppecher, and Bouchitt\'e \cite{Milton:2009:MVP} and Milton and Willis \cite{Milton:2010:MVP},
we can rewrite this consititutive law in the inclusion phase as
\beq \underbrace{\begin{pmatrix} -\Imag (\CF^s_2(\Bx))\\ \Imag (\CG^s_2(\Bx))\end{pmatrix}}_{\BJ_2^0(\Bx)}
=\underbrace{\begin{pmatrix} [\BZ_1'']^{-1} & -[\BZ_1'']^{-1}\BZ_1' \\ -\BZ_1'[\BZ_1'']^{-1} & \BZ_1''+ \BZ_1'[\BZ_1'']^{-1}\BZ_1'\end{pmatrix}}_{\BCL_1}
\underbrace{\begin{pmatrix} \Real (\CG^s_2(\Bx)) \\ \Real (\CF^s_2(\Bx))\end{pmatrix}}_{\BE_2^0(\Bx)},
\eeq{5.2}
where the matrix $\BCL_1$ is now positive definite. In the matrix phase a relation like \eq{5.2} does not hold as $\BZ_0$ is real. However, what enters
the variational principle is $\BE_2^0(\Bx)\cdot\BJ_2^0=\BE_2^0(\Bx)\cdot\BCL\BE_2^0$. This will remain finite (and in fact approaches zero)
as the imaginary part of $\BZ$ tends to zero if
the fields $\BJ_2^0(\Bx)$ and $\BE_2^0(\Bx)$ defined in \eq{5.2} are required to have components satisfying 
\beq \Imag (\CG^s_2(\Bx))=\BZ_0\Imag (\CF^s_2(\Bx)),\quad \Real (\CG^s_2(\Bx))=\BZ_0\Real (\CF^s_2(\Bx)), \eeq{5.2a}
as implied by \eq{3.9}, where $\BZ_0$ is real. Thus we have
\beq \BE_2^0(\Bx)\cdot\BJ_2^0(\Bx)=-\Real (\CG^s_2(\Bx))\cdot\Imag (\CF^s_2(\Bx))+\Real (\CF^s_2(\Bx))\cdot\Imag (\CG^s_2(\Bx))=0 \eeq{5.2b}
for all $\Bx$ in the matrix. 
Similarly, on the subspace $\CS_3$ where $\BL$ maps a field $[0,\{\Bq_{\CF}\}_1,\{\Bq_{\CF}\}_2]$ to $[0,\{\Bq_{\CG}\}_1,\{\Bq_{\CG}\}_2]$ 
the constitutive law \eq{3.9a},
$\Bq_{\CG}=\frac{i\Go}{\kappa_0}\Bq_{\CF}$ can be rewritten, analogously to \eq{5.2}, as 
\beq \underbrace{\begin{pmatrix} -\Imag(\Bq_{\CF}(\Bn) \\ \Imag (\Bq_{\CG}(\Bn))\end{pmatrix}}_{\Bt(\Bn)}=\underbrace{\begin{pmatrix} \kappa_0/\Go  & 0 \\ 0 & \Go/\kappa_0 \end{pmatrix}}_{\BCL_3}
\underbrace{\begin{pmatrix} \Real(\Bq_{\CG}(\Bn))\\ \Real (\Bq_{\CF}(\Bn))\end{pmatrix}}_{\Bs(\Bn)},
\eeq{5.3}
where $\BCL_3$ is clearly positive definite.
Now suppose we have a real trial pressure field $\underline{P}(\Bx)$ and a purely imaginary trial velocity field $\underline{\Bv}(\Bx)$, such that the associated real fields
\beqa \underline{\CF}^0(\Bx)& = & \begin{pmatrix}\Grad \underline{P}(\Bx) \\ \underline{P}(\Bx) \end{pmatrix}, \nonum
     \underline{\CG}^0(\Bx) & = & \begin{pmatrix}-i\underline{\Bv} \\ -i\Div\underline{\Bv} \end{pmatrix}, 
\eeqa{5.4}
have the asymptotic forms
\beqa  
\underline{\CF}^0(\Bx) & = &\frac{e^{ik_0|\Bx|}}{|\Bx|}\left\{\frac{\underline{P}_\infty({\widehat{\Bx}})}{2}\begin{pmatrix} ik_0\widehat{\Bx} \\ 1 \end{pmatrix}
 +\mathcal{O}\left(\frac{1}{|\Bx|}\right)\right\}+\frac{e^{-ik_0|\Bx|}}{|\Bx|}\left\{\frac{\overline{\underline{P}_\infty({\widehat{\Bx}})}}{2}\begin{pmatrix} -ik_0\widehat{\Bx} \\ 1 \end{pmatrix}+\mathcal{O}\left(\frac{1}{|\Bx|}\right)\right\},\nonum
\underline{\CG}^0(\Bx) & = & \frac{e^{ik_0|\Bx|}}{|\Bx|}\left\{\frac{\underline{V}_\infty({\widehat{\Bx}})}{2}\begin{pmatrix} -i\widehat{\Bx}/k_0 \\ 1 \end{pmatrix}
+\mathcal{O}\left(\frac{1}{|\Bx|}\right)\right\}+\frac{e^{-ik_0|\Bx|}}{|\Bx|}\left\{\frac{\overline{\underline{V}_\infty({\widehat{\Bx}})}}{2}\begin{pmatrix} i\widehat{\Bx}/k_0 \\ 1 \end{pmatrix}+\mathcal{O}\left(\frac{1}{|\Bx|}\right)\right\},
\eeqa{5.5}
for some choice of complex valued functions $\underline{P}_\infty(\Bn)$ and $\underline{V}_\infty(\Bn)$,
and are such that in the matrix (outside the inclusion)
\beq \underline{\CG}^0(\Bx)=\BZ_0\underline{\CF}^0(\Bx), \eeq{5.6}
so that the trial fields satisfy \eq{5.2a}. Thus in the matrix the trial fields are required to be solutions to the acoustic wave equation. 
From the asymptotic form \eq{5.5} and \eq{3.5fba} we see that the accompanying auxillary fields are
\beq  \Bq_{\underline{\CF}}(\Bn)=[\Imag(\underline{P}_\infty(\Bn)), \Real(\underline{P}_\infty(\Bn))],~~{\rm and}~~ \Bq_{\underline{\CG}}(\Bn)=[\Real(\underline{V}_\infty(\Bn)), -\Imag(\underline{V}_\infty(\Bn))].
\eeq{5.6a}
The fields $\underline{\CF}^0(\Bx)$ and $\underline{\CG}^0(\Bx)$ can be expressed as 
\beq \underline{\CF}^0=\underline{\CF}_1^0+\underline{\CF}_2^0, \quad \underline{\CG}^0=\underline{\CG}_1^0+\underline{\CG}_2^0,\quad{\rm with}\quad
\underline{\CF}_1^0,\underline{\CG}_1^0\in\CV^0,\quad \underline{\CF}_2^0,\underline{\CG}_2^0\in\CH^0.
\eeq{5.7}
So if we define
\beq \underline{\BE}_2^0(\Bx)=\begin{pmatrix} \underline{\CG}_2^0(\Bx) \\ \underline{\CF}_2^0(\Bx) \end{pmatrix},
\eeq{5.7a}
it follows from \eq{5.6} that $\underline{\CG}_2^0(\Bx)=\BZ_0\underline{\CF}_2^0(\Bx)$ and then \eq{5.2} and \eq{5.2a} imply
\beq \underline{\BE}_2^0(\Bx)\cdot\BJ_2^0(\Bx)=0, \eeq{5.7b}
for all $\Bx$ in the matrix.

Suppose now we prescribe
\beq \Real (\CF^s_1)=\underline{\CF}_1^0,\quad \Real (\CG^s_1)=\underline{\CG}_1^0, \eeq{5.9}
and let $\CF^s=\CF^s_1+\CF^s_2$ and $\CG^s=\CG^s_1+\CG^s_2$ be the associated solutions of the $Y$-problem. Then, as shown in appendix A,
we have the variational inequality,
\beqa &~& \int_{\GO}\underline{\BE}_2^0(\Bx)\cdot\BCL_1\underline{\BE}_2^0(\Bx)\,d\Bx+\int_{|\Bn|=1}\underline{\Bs}(\Bn)\cdot\BCL_3\underline{\Bs}(\Bn)\,dS 
\geq \quad (\Imag(\CF^s_1),\Real (\CG^s_1))-(\Real (\CF^s_1),\Imag(\CG^s_1))
=- (\BJ_1^0,\BE_1^0), \nonum &~&
\eeqa{5.17}
where 
\beq \underline{\Bs}(\Bn)=\bpm \Bq_{\underline{\CG}}(\Bn) \\ \Bq_{\underline{\CF}}(\Bn) \epm, \quad
\BJ_1^0(\Bx)=\begin{pmatrix} -\Imag (\CF^s_1(\Bx))\\ \Imag (\CG^s_1(\Bx))\end{pmatrix},\quad
\BE_1^0=\begin{pmatrix} \Real (\CG^s_1(\Bx)) \\ \Real (\CF^s_1(\Bx))\end{pmatrix}.
\eeq{5.17a}
From the definition \eq{3.10a} of the $Y$-operator, $\BY_*^0$, we have $\CG_1^s=-\BY_*^0\CF_1^s$ and this relation
can then be manipulated into the form
\beq \underbrace{\begin{pmatrix} -\Imag (\CF^s_1)\\ \Imag (\CG^s_1)\end{pmatrix}}_{\BJ_1}
=-\BCY
\underbrace{\begin{pmatrix} \Real (\CG^s_1) \\ \Real (\CF^s_1)\end{pmatrix}}_{\BE_1},
\eeq{5.24}
which defines the associated operator $\BCY$, and the fields $\BJ_1$ and $\BE_1$. Then
the right hand sides of \eq{5.17a} can then be identified with the quadratic form associated with $\BCY$:
\beq  (\Imag(\CF^s_1),\Real (\CG^s_1))-(\Real (\CF^s_1),\Imag(\CG^s_1))=-(\BE_1,\BJ_1)=(\BE_1,\BCY\BE_1).
\eeq{5.25}
Consequently we have the variational principle
\beq (\BE_1,\BCY\BE_1)=\min_{\underline{\BE}_2^0} \int_{\GO}\underline{\BE}_2^0(\Bx)\cdot\BCL_1\underline{\BE}_2^0(\Bx)\,d\Bx+\int_{|\Bn|=1}\underline{\Bs}(\Bn)\cdot\BCL_3\underline{\Bs}(\Bn)\,dS 
\eeq{5.25a}
where the minimum is over all fields $\underline{\BE}_2^0$ such that $\underline{\BE}^0=\BE_1+\underline{\BE}_2^0$ is of the form
\beq \underline{\BE}^0=\underline{\BE}^0(\Bx)=\begin{pmatrix} \underline{\CG}^0(\Bx) \\ \underline{\CF}^0(\Bx) \end{pmatrix},
\eeq{5.25b}
with $\underline{\CG}^0(\Bx)$ and $\underline{\CF}^0(\Bx)$ being of the form \eq{5.4} for some real $\underline{P}(\Bx)$ and a purely imaginary vector field $\underline{\Bv}(\Bx)$.
Additionally, the constitutive relation \eq{5.6} must hold in the matrix. As the right hand side of \eq{5.25a} is non-negative, we deduce that $\BCY$ is a positive semidefinite operator.

Expressing $\underline{\CF}_2^0(\Bx)$ and $\underline{\CG}_2^0(\Bx)$ in terms of their component fields,
\beq \underline{\CF}_2^0(\Bx))=\bpm \BF(\Bx) \\ f(\Bx) \epm, \quad \underline{\CG}_2^0(\Bx))=\bpm \BG(\Bx) \\ g(\Bx) \epm, 
\eeq{5.18}
the inequality \eq{5.17} takes the equivalent form
\beqa &~& \int_\GO \bpm \BG(\Bx) \\ \BF(\Bx) \epm\cdot\BR\bpm \BG(\Bx) \\ \BF(\Bx) \epm +\bpm g(\Bx) \\ f(\Bx) \epm\cdot\BH\bpm g(\Bx) \\ f(\Bx) \epm\,d\Bx
 +  \int_{|\Bn|=1}\kappa_0|\underline{V}_\infty(\Bn)|^2/\Go+\Go|\underline{P}_\infty(\Bn)|^2/\kappa_0\,dS \nonum
&~& \quad\quad \geq  (\Imag(\CF^s_1),\Real (\CG^s_1))-(\Real (\CF^s_1),\Imag(\CG^s_1)).
\eeqa{5.19}
in which
\beqa \BR & = & \bpm\Go(\Br'')^{-1} & -(\Br'')^{-1}\Br' \\ 
                   -\Br'(\Br'')^{-1} & [\Br''+\Br'(\Br'')^{-1}\Br']/\Go \epm, \nonum
\BH & = & \bpm (\Go h'')^{-1} & -(h'')^{-1}h' \\
                   -h'(h'')^{-1} & \Go[h''+h'(h'')^{-1}h'] \epm,
\eeqa{5.20}
and $\Br=-\BGr_1^{-1}$, and $h=1/\Gk_1$. When $\Br''$ is very small we have
\beq \bpm \BG(\Bx) \\ \BF(\Bx) \epm\cdot\BR\bpm \BG(\Bx) \\ \BF(\Bx) \epm
\approx (\Go\BG(\Bx)-\Br'\BF(\Bx))\cdot(\Go\Br'')^{-1}(\Go\BG(\Bx)-\Br'\BF(\Bx),
\eeq{5.21}
so if this is to remain finite in the limit $\Br''\to 0$ (i.e., when $\BGr_1$ is real) we need to choose the trial fields so that
\beq \BF(\Bx)=-\Go\BGr_1\BG(\Bx)~~~{\rm for~all~}\Bx\in\GO. 
\eeq{5.22}
Then, taking the limit $\Br''\to 0$, the variational inequality \eq{5.19} reduces to
\beq  \int_\GO \bpm g(\Bx) \\ f(\Bx) \epm\cdot\BH\bpm g(\Bx) \\ f(\Bx) \epm\,d\Bx
+\int_{|\Bn|=1}\kappa_0|\underline{V}_\infty(\Bn)|^2/\Go+\Go|\underline{P}_\infty(\Bn)|^2/\kappa_0\,dS
\geq  (\Imag(\CF^s_1),\Real (\CG^s_1))-(\Real (\CF^s_1),\Imag(\CG^s_1)).
\eeq{5.23}

\section{The link between the power absorbed and scattered by the inclusion and $\Imag(\BY_*^0)$.}
\setcounter{equation}{0}
The imaginary part of the quadratic form associated with $\BY_*^0$ has a physical interpretation
in terms of the power absorbed and scattered by the inclusion. In elastodynamics 
the power absorption by a body $\GO$, having a possibly complex density $\BGr_1=\BGr_1'+i\BGr_1''$
(with real and imaginary parts $\BGr_1'$ and $\BGr_1''$), is given by
formula (2.5) in \cite{Milton:2010:MVP} and (taking into account our choice of $e^{-i\Go t}$ for
the time dependence, rather than $e^{i\Go t}$)
can be written as
\beq A=\frac{1}{2}\int_{\GO}\Go\overline{\Bv^0}\cdot\BGr_1''\Bv^0+\Real(\overline{-i\Go\Be^0}\cdot\BGs^0)\,d\Bx
\eeq{6.10a}
where $\Bv^0=-i\Go\Bu^0$ is the complex velocity field, $\Bu^0$ is the complex displacement field,
$\Be^0=[\Grad\Bu^0+(\Grad\Bu^0)^T]/2$ is the strain, and its time derivative $-i\Go\Be^0=[\Grad\Bv^0+(\Grad\Bv^0)^T]/2$ is the
strain rate, and $\BGs^0$ is the stress. In a fluid one has $\BGs^0=-P^0\BI$ where $P^0(\Bx)$ is the pressure, 
and hence the above expression reduces to
\beqa  A & = & \frac{1}{2}\int_{\GO}\Go\Imag(\overline{\Bv^0}\cdot\BGr_1\Bv^0)-\Real(\overline{\Div\Bv^0}P^0)\,d\Bx \nonum
& = & \frac{1}{2}\int_{\GO}\Imag(\overline{i\Bv^0}\cdot\Grad P^0)+\Imag(\overline{i\Div\Bv}P^0)\,d\Bx \nonum
& = & \frac{1}{2}\int_{\GO}\Imag(-i\Bv\cdot\overline{\Grad P^0})\}+\Imag(-i\Div\Bv\overline{P^0})\,d\Bx \nonum
& = & \frac{1}{2}\int_{\GO}\Imag(\overline{\CF^0}\cdot\CG^0)\,d\Bx \nonum
& = & \frac{1}{2}\int_{\GO}\Real(\CF)\cdot\Imag(\CG^0)-\Imag(\CF^0)\cdot\Real(\CG),
\eeqa{6.10b}
where 
\beq \CG^0=\bpm -i\Bv^0 \\ -i\Div\Bv^0 \epm, \quad \CF^0=\bpm \Grad P^0 \\ P^0 \epm.
\eeq{6.10c}
Thus by taking the imaginary part of the key identity \eq{3.5faa} we see that twice the imaginary part of the left hand side can be identified 
with the time-averaged power absorbed in the ball $B_r$ and consequently, by conservation of energy,
$$ \frac{1}{2}\Imag\int_{\Md B_{r}}-i\overline{P^0(\Bx)}\Bn\cdot\Bv^0(\Bx)\,dS $$
can be identified with the time-averaged power flowing inwards through the boundary $\Md B_{r}$. Hence in the identity
\beq \frac{1}{2}\Imag(\CF_1^s,\BY_*^0\CF_1^s)=\frac{1}{2}\Imag\int_{\GO}(\overline{\CF^s_2}\cdot\BZ_1\CF^s_2)\,d\Bx
+\frac{1}{2}\Imag\int_{\Md B_{r}}i\overline{P^s(\Bx)}\Bn\cdot\Bv^s(\Bx)\,dS,
\eeq{6.10d}
implied by \eq{3.11a} and \eq{3.5fab} (with $P^i_\infty=0$ and $V^i_\infty=0$), we see that the first term on the right
can be identified with the time-averaged power absorbed by the field $\CF^s_2$ in the inclusion, while the second term on the right
can be identified with the time-averaged power radiated to infinity by the scattered field.

The total time-averaged power absorbed by the inclusion has contributions both from the field $\CF^s_2$ and from the fields $\CF^a+\CF_1^s$, 
and is given by
\beq \frac{1}{2}\Imag\int_{\GO}(\overline{(\CF^a+\CF_1^s+\CF^s_2)}\cdot\BZ_1(\CF^a+\CF_1^s+\CF^s_2))\,d\Bx
=\frac{1}{2}\Imag\int_{\GO}(\overline{(\CF^a+\CF_1^s)}\cdot\BZ_1(\CF^a+\CF_1^s))\,d\Bx
+\frac{1}{2}\Imag\int_{\GO}(\overline{\CF^s_2}\cdot\BZ_1\CF^s_2)\,d\Bx,
\eeq{6.10e}
where in obtaining this last identity we have used the orthgonality of the spaces $\CV^0$ and $\CH^0$. The last term in \eq{6.10e}
is that which enters \eq{6.10d}.

The time averaged extinction power, being the sum of the total absorbed power and scattered power should be $W$, where
from \eq{6.10d} and \eq{6.10e},
\beqa 2W  & = &  \Imag(\CF^a+\CF_1^s,\Gc\BZ_1(\CF^a+\CF_1^s))+\Imag(\CF_1^s,\BY_*^0\CF_1^s) \nonum
 &=& (\CF^a+\CF_1^s,\Gc\Imag(\BZ_1)(\CF^a+\CF_1^s))+(\CF_1^s,\Imag(\BY_*^0)\CF_1^s)
\eeqa{6.10f}
This provides the desired link between $W$ and $\Imag(\BY_*^0)$.
Further manipulations, carried out in Appendix B, provide alternative expressions for $W$, namely
\beq 2W= \Imag\int_{B_{r}}\overline{p^a}{\CF^a}'\cdot(\CG^s-\BZ_0\CF^s)\,d\Bx, \eeq{6.12}
which is similar to the form of the optical theorem given in \cite{Lytle:2005:GOT,Hashemi:2012:DBP,Miller:2015:FLO}, and
\beq W  =  2k_0\pi\Imag[\overline{p^a} P^s_\infty(\Bk_0/k_0)]/(\Go\Gr_0),
\eeq{6.14}
which is the well-known form of the optical theorem \cite{Newton:1976:OTB,Dassios:2000:LFS}
for acoustic scattering.

\section{Variational principles for the backwards scattering amplitude}
\setcounter{equation}{0}
For $\Bx$ in the inclusion phase, the constitutive law implies $\CG^a+\CG_1^s=\BZ_1(\CF^a+\CF_1^s)$ which analogously
to \eq{5.1} and \eq{5.2} can be manipulated into the form
\beq  \BJ^a+\BJ_1=\BCL_1(\BE^a+\BE_1),\quad {\rm where}~ \BE^a=\begin{pmatrix} \Real (\CG^a(\Bx)) \\ \Real (\CF^a(\Bx))\end{pmatrix},\quad
\BJ^a=\begin{pmatrix} -\Imag (\CF^a(\Bx)) \\ \Imag (\CG^a(\Bx))\end{pmatrix}
\eeq{7.5a}
and $\BJ_1$ and $\BE_1$ are defined in \eq{5.24}. Thus the formula \eq{6.10f} for the extinction power can be rewritten as 
\beqa 2W  & = &  \Imag(\CF^a+\CF_1^s,\Gc(\CG^a+\CG_1^s))-\Imag(\CF_1^s,\CG_1^s) \nonum
& = & (\BE^a+\BE_1,\Gc(\BJ^a+\BJ_1))-(\BE_1,\BJ_1)\nonum
& = & (\BE^a+\BE_1,\Gc\BCL_1(\BE^a+\BE_1))+(\BE_1,\BCY\BE_1)
\eeqa{7.5b}
As $\BCL_1$ and $\BCY$ are positive semidefinite operators, this formula suggests that a variational principle might be obtained from a consideration
of the non-negativity of the quadratic form
\beqa &~& (\BE^a+\underline{\BE}_1-\BE^a-\BE_1,\BCL_1(\BE^a+\underline{\BE}_1-\BE^a-\BE_1))
+(\underline{\BE}_1-\BE_1,\BCY(\underline{\BE}_1-\BE_1))\nonum
& ~ & = (\BE^a+\underline{\BE}_1,\BCL_1(\BE^a+\underline{\BE}_1))+(\BE^a+\BE_1,\BCL_1(\BE^a+\BE_1))
+(\underline{\BE}_1\BCY\underline{\BE}_1)+(\BE_1,\BCY\BE_1)\nonum
& ~& \quad -2(\BE^a+\underline{\BE}_1,\BCL_1(\BE^a+\BE_1))-2(\underline{\BE}_1,\BCY\BE_1),
\eeqa{7.5}

The sum of the last two terms in \eq{7.5}, each of which involves both $\BE_1$ and $\underline{\BE}_1$, can be replaced by the expression
\beqa &~& -2(\BE^a+\underline{\BE}_1,\BCL_1(\BE^a+\BE_1))-2(\underline{\BE}_1,\BCY\BE_1) \nonum
&~&\quad=-2(\BE^a+\underline{\BE}_1,\BJ^a+\BJ_1)+2(\underline{\BE}_1,\BJ_1)\nonum
&~&\quad=-2(\BE^a,\BJ^a+\BJ_1)-2(\underline{\BE}_1,\BJ^a)\nonum
&~&\quad=-2(\BE^a+\BE_1,\BJ^a+\BJ_1)+2(\BE_1,\BJ_1)+2(\BE_1,\BJ^a)-2(\underline{\BE}_1,\BJ^a)\nonum
&~&\quad =-2(\BE^a+\BE_1,\BCL_1(\BE^a+\BE_1))-2(\BE_1,\BCY\BE_1)+2(\BE_1,\BJ^a)-2(\underline{\BE}_1,\BJ^a).
\eeqa{7.6}
By substituting this back in \eq{7.5} one sees that one has the variational principle
\beqa (\BE^a+\BE_1,\BCL_1(\BE^a+\BE_1))+(\BE_1,\BCY\BE_1)-2(\BE_1,\BJ^a)
=\min_{\underline{\BE}_1}(\BE^a+\underline{\BE}_1,\BCL_1(\BE^a+\underline{\BE}_1))+(\underline{\BE}_1\BCY\underline{\BE}_1)
-2(\underline{\BE}_1,\BJ^a).
\eeqa{7.7}
The variational principle derived in Section \sect{varprin} can then be substituted into this expression and we obtain
\beqa 2W-2(\BE_1,\BJ^a)& = & \min_{\underline{\BP},\underline{\Bv}}
(\BE^a+\underline{\BE}_1,\BCL_1(\BE^a+\underline{\BE}_1))+(\underline{\BE}_2,\BCL_1\underline{\BE}_2))-2(\underline{\BE}_1,\BJ^a)
 +\int_{|\Bn|=1}\kappa_0|\underline{V}_\infty(\Bn)|^2/\Go+\Go|\underline{P}_\infty(\Bn)|^2/\kappa_0\,dS\nonum
& = &\min_{\underline{\BP},\underline{\Bv}}
(\BE^a+\underline{\BE},\BCL_1(\BE^a+\underline{\BE}))-2(\underline{\BE},\Gc\BJ^a)
 +\int_{|\Bn|=1}\kappa_0|\underline{V}_\infty(\Bn)|^2/\Go+\Go|\underline{P}_\infty(\Bn)|^2/\kappa_0\,dS.
\eeqa{7.8}
where here $\underline{\BP}$ is a real trial pressure field, and $\underline{\Bv}$ is a purely imaginary trial velocity field, and the real field
$\underline{\BE}(\Bx)$ is given in terms of them through the equations
\beqa \underline{\BE} & = &\underline{\BE}_1+\underline{\BE}_2=\begin{pmatrix} \underline{\CG}^0(\Bx)) \\ \underline{\CF}^0(\Bx))\end{pmatrix},\nonum
\underline{\CF}^0(\Bx)& = & \begin{pmatrix}\Grad \underline{P}(\Bx) \\ \underline{P}(\Bx) \end{pmatrix}\in\CE^0, \quad
       \underline{\CG}^0(\Bx)  =  \begin{pmatrix}-i\underline{\Bv} \\ -i\Div\underline{\Bv} \end{pmatrix}\in\CJ^0, 
\eeqa{7.9}
where $\CE^0$ and $\CJ^0$ consist of all fields of the form \eq{3.5ba} and \eq{3.5bb}, respectively. 

This variational principle has the advantage that the quantity on the right hand side of \eq{7.8} is easy to numerically compute
for a given choice of $\underline{\BE}$:  it is not necessary to determine the individual component fields $\underline{\BE}_1$ and
$\underline{\BE}_2=\underline{\BE}-\underline{\BE}_1$. To obtain a physical interpretation for the quantity $-2(\BE_1,\BJ^a)$ appearing 
on the left hand side of \eq{7.8} note that
\beqa -2(\BE_1,\BJ^a) & = & 2(\Real \CG_1, \Imag \CF^a)+2(\Real \CF_1, -\Imag \CG^a) \nonum
& = & 2(\Real(\CG_1-\BZ_0\CF_1),\Imag \CF^a) \nonum
& = & 2(\Real(\CG^s-\BZ_0\CF^s),\Imag \CF^a) \nonum
& = & 2\int_{B_r}\Imag(\CF^a)\cdot[\Real(\CG^s-\BZ_0\CF^s)]\,d\Bx \nonum
& = &  2\int_{B_r}\Imag(\CF^a)\cdot\Real\CG^s-\Imag(\CG^a)\cdot\Real\CF^s\,d\Bx \nonum
& = & 2\int_{\Md B_r}\Imag(P^a)\Imag(\Bn\cdot\Bv^s)+\Real(\Bn\cdot\Bv^a)\Real(P^s)\,dS.
\eeqa{7.10} 
Using the asymptotic forms of the fields as $r\to\infty$ we get
\beqa
-2(\BE_1,\BJ^a) & = & 2\int_{\Md B_r}\frac{k_0\Imag(p^ae^{i\Bk_0\cdot\Bx})\Imag(P^s_\infty(\widehat{\Bx})e^{ik_0r})}{r\Go\Gr_0}
+\frac{(\Bn\cdot\Bk_0)\Real(p^ae^{i\Bk_0\cdot\Bx})\Real(P^s_\infty(\widehat{\Bx})e^{ik_0r})}{r\Go\Gr_0}.
\eeqa{7.11}
Choosing our coodinates so that the positive $x_1$ axis points in the direction of $\Bk_0$, i.e. so that
$\Bk_0\cdot\Bx=k_0x_1=k_0rt$ and $\Bn\cdot\Bk_0=k_0t$ where $t=x_1/r$, and making the substitutions
\beqa \Imag(p^ae^{i\Bk_0\cdot\Bx}) & = & (-ip^ae^{ik_0rt}+i\overline{p^a}e^{-ik_0rt})/2,\quad
\Real(p^ae^{i\Bk_0\cdot\Bx})=(p^ae^{ik_0rt}+\overline{p^a}e^{-ik_0rt})/2,\nonum
\Imag(P^s_\infty(\widehat{\Bx})e^{ik_0r})& = & (-iP^s_\infty(\widehat{\Bx})e^{ik_0r}+i\overline{P^s_\infty(\widehat{\Bx})}e^{-ik_0r})/2,\quad
\Real(P^s_\infty(\widehat{\Bx})e^{ik_0r})=(P^s_\infty(\widehat{\Bx})e^{ik_0r}+\overline{P^s_\infty(\widehat{\Bx})}e^{-ik_0r})/2 \nonum
&~&
\eeqa{7.12}
we are left with $-2(\BE_1,\BJ^a)$ being the sum of the two integrals
\beqa \frac{1}{2}\int_{\Md B_r}\frac{k_0(t-1)p^a P^s_\infty(\widehat{\Bx})e^{ik_0r(1+t)}}{r\Go\Gr_0}
=\frac{p^a}{2\Go\Gr_0}\int_{-1}^1rk_0(t-1)p_\infty(t)e^{ik_0r(1+t)}\,dt, \nonum
\frac{1}{2}\int_{\Md B_r}\frac{k_0(t+1)\overline{p^a}P^s_\infty(\widehat{\Bx})e^{ik_0r(1-t)}}{r\Go\Gr_0}
=\frac{\overline{p^a}}{2\Go\Gr_0}\int_{-1}^1rk_0(t+1)p_\infty(t)e^{ik_0r(1-t)}\,dt,
\eeqa{7.13}
and their complex conjugates, in which $p_\infty(t)$ is defined by \eq{4.20}.
The integrals are of the same form as the integral $I_2$ in \eq{4.21}, 
with appropriate choices of $f(t)$ and $g(t)$. Using the formula \eq{4.24} we can evaluate them in the limit
$r\to\infty$ and they equal respectively
$$ -ip^a2\pi k_0P^s_\infty(-\Bk_0/k_0)/(\Go\Gr_0),\quad {\rm and}~~i\overline{p^a}2\pi k_0P^s_\infty(\Bk_0/k_0)/(\Go\Gr_0). $$
Adding them, and then adding the total to its complex conjugate gives
\beqa -2(\BE_1,\BJ^a) & = & 4\pi k_0\Imag(p^a P^s_\infty(-\Bk_0/k_0))/(\Go\Gr_0)
-4\pi k_0\Imag(\overline{p^a} P^s_\infty(\Bk_0/k_0))/(\Go\Gr_0) \nonum
& = & 4\pi k_0\Imag(p^a P^s_\infty(-\Bk_0/k_0))/(\Go\Gr_0)-2W, \eeqa{7.14}
where we have used the expression \eq{6.14} for W given by the optical theorem.
Thus we have the variational principle:
\beq 4\pi k_0\Imag(p^a P^s_\infty(-\Bk_0/k_0))/(\Go\Gr_0)
 = \min_{\underline{\BP},\underline{\Bv}}
(\BE^a+\underline{\BE},\BCL_1(\BE^a+\underline{\BE}))-2(\underline{\BE},\Gc\BJ^a)
 +\int_{|\Bn|=1}\kappa_0|\underline{V}_\infty(\Bn)|^2/\Go+\Go|\underline{P}_\infty(\Bn)|^2/\kappa_0\,dS.
\eeq{7.15}
It is interesting that this variational principle, with some choice of trial fields $\underline{\BP}$ and $\underline{\Bv}$, does not
give a desired bound on $W$, or equivalently on the forward scattering amplitude, but rather bounds the {\it backwards} scattering amplitude
$P^s_\infty(-\Bk_0/k_0)$.

We note that the physical pressure field associated with the incoming wave is $\Real(P^a(\Bx)e^{-i\Go t})$ where $t$ is the time.
Accordingly, if we shift our origin of time, by replacing $t$ with $t-t_0$, the physical pressure field associated with the incoming wave is
$\Real(\widetilde{P}^a(\Bx)e^{-i\Go t})$ where 
\beq \widetilde{P}^a(\Bx)=P^a(\Bx)e^{i\Go t_0}=\widetilde{p}^ae^{i\Bk_0\cdot\Bx}~~~{\rm where}~~\widetilde{p}^a=p^ae^{i\Go t_0}.
\eeq{7.16}
The associated scattered pressure field is then
\beq \widetilde{P}^s(\Bx)=P^s(\Bx)e^{i\Go t_0},\quad {\rm with}~~\widetilde{P}^s_\infty(\widehat{\Bx})=P^s_\infty(\widehat{\Bx})e^{i\Go t_0}.
\eeq{7.17}
Consequently, with $p^a$ and $P^s_\infty(\widehat{\Bx})$ replaced by $\widetilde{p}^a$ and $\widetilde{P}^s_\infty(\widehat{\Bx})$,
the variational principle \eq{7.15} becomes
\beq 4\pi k_0\Imag(e^{2i\Go t_0}p^a P^s_\infty(-\Bk_0/k_0))/(\Go\Gr_0)
 = \min_{\underline{\BP},\underline{\Bv}}
(\widetilde{\BE}^a+\underline{\BE},\BCL_1(\widetilde{\BE}^a+\underline{\BE}))-2(\underline{\BE},\Gc\widetilde{\BJ}^a)
 +\int_{|\Bn|=1}\kappa_0|\underline{V}_\infty(\Bn)|^2/\Go+\Go|\underline{P}_\infty(\Bn)|^2/\kappa_0\,dS,
\eeq{7.18}
where
\beq \widetilde{\BE}^a(\Bx)=\begin{pmatrix} \Real (e^{i\Go t_0}\CG^a(\Bx)) \\ \Real (e^{i\Go t_0}\CF^a(\Bx))\end{pmatrix},\quad
\widetilde{\BJ}^a(\Bx)=\begin{pmatrix} -\Imag (e^{i\Go t_0}\CF^s_1(\Bx))\\ \Imag (e^{i\Go t_0}\CG^s_1(\Bx))\end{pmatrix}.
\eeq{7.19}
Thus by varying $t_0$, and appropriately changing the trial fields, one get bounds that ``wrap around'' the possible complex
values of the backwards scattering amplitude $P^s_\infty(-\Bk_0/k_0)$. By choosing the origin of time appropriately we can assume that $p^a$ is real and positive.
Then, for example, \eq{7.15} provides an upper bound on $\Imag(P^s_\infty(-\Bk_0/k_0))$, while \eq{7.18} with $t_0$ chosen so that
$e^{2i\Go t_0}=-1$ provides a lower bound on $\Imag(P^s_\infty(-\Bk_0/k_0))$. Similarly \eq{7.18}, with $t_0$ chosen so that
$e^{2i\Go t_0}=-i$ or $e^{2i\Go t_0}=i$, gives us upper and lower bounds on $\Real(P^s_\infty(-\Bk_0/k_0))$.

The simplest choice for the trial field $\underline{\BE}$ is of course $\underline{\BE}=0$ and 
(still assuming the origin of time has been chosen so $p^a$ is real and positive) this gives
\beqa  &~ & 4\pi k_0\Imag(e^{2i\Go t_0}p^a P^s_\infty(-\Bk_0/k_0))/(\Go\Gr_0)\leq (\BE^a,\BCL_1\BE^a) \nonum
& ~ & \leq \int_\GO\begin{pmatrix} \BZ_0\Real (e^{i\Go t_0}\CF^a(\Bx)) \\ \Real (e^{i\Go t_0}\CF^a(\Bx))\end{pmatrix}\cdot
\begin{pmatrix} [\BZ_1'']^{-1} & -[\BZ_1'']^{-1}\BZ_1' \\ -\BZ_1'[\BZ_1'']^{-1} & \BZ_1''+ \BZ_1'[\BZ_1'']^{-1}\BZ_1'\end{pmatrix}
\begin{pmatrix} \BZ_0\Real (e^{i\Go t_0}\CF^a(\Bx)) \\ \Real (e^{i\Go t_0}\CF^a(\Bx))\end{pmatrix}\,d\Bx \nonum
& ~ & \leq \int_\GO \Real (e^{i\Go t_0}\CF^a(\Bx))\cdot[\BZ_1''+(\BZ_1'-\BZ_0)[\BZ_1'']^{-1}(\BZ_1'-\BZ_0)]\Real (e^{i\Go t_0}\CF^a(\Bx))\,d\Bx \nonum
 & ~ & \leq \int_\GO \Real(e^{i\Go t_0}\Grad P^a)\cdot [\Br''+(\Br'-\Br_0)(\Br'')^{-1}(\Br'-\Br_0)]\Real((e^{i\Go t_0}\Grad P^a)/\Go \nonum
& ~&\quad +\Go\Real(e^{i\Go t_0}P^a)[h_1''+(h_1'-h_0)(h_1'')^{-1}(h_1'-h_0)]\Real(e^{i\Go t_0}P^a)\,d\Bx \nonum
& ~ & \leq \frac{(p^a)^2\Bk_0\cdot[\Br_1''+(\Br_1'-\Br_0)(\Br_1'')^{-1}(\Br_1'-\Br_0)]\Bk_0}{\Go}\int_\GO\{\sin[(\Bk_0\cdot\Bx)+\Go t_0]\}^2\,d\Bx \nonum
& ~ & +\Go(p^a)^2[h_1''+(h_1'-h_0)^2(h_1'')^{-1}]\int_\GO\{\cos[(\Bk_0\cdot\Bx)+\Go t_0]\}^2\,d\Bx,
\eeqa{8.3}
and $\Br_1=-\BGr_1^{-1}=\Br_1'+i\Br_1''$, and $h_1=1/\Gk_1=h_1'+ih_1''$. If both the inclusion phase and the matrix phase are isotropic, so that
$\Br_1'=r_1'\BI$ and $\Br_1''=r_1''\BI$ then the bound becomes
\beqa 4\pi k_0\Imag(e^{2i\Go t_0}p^a P^s_\infty(-\Bk_0/k_0))/(\Go\Gr_0)
& \leq &  \frac{k_0^2(p^a)^2[r_1''+(r_1'-r_0)^2/r_1'']}{\Go}\int_\GO\{\sin[(\Bk_0\cdot\Bx)+\Go t_0]\}^2\,d\Bx \nonum
& ~ & \quad +\Go(p^a)^2[h_1''+(h_1'-h_0)^2/h_1'']\int_\GO\{\cos[(\Bk_0\cdot\Bx)+\Go t_0]\}^2\,d\Bx.
\eeqa{8.3a}
We can express the bound directly in terms of the real and imaginary parts of the complex density $\Gr_1=\Gr_1'+i\Gr_1''$ and complex bulk modulus 
$\Gk_1=\Gk_1'+i\Gk_1''$
using the identities
\beq [r_1''+(r_1'-r_0)^2/r_1'']=[\Gr_1''+(\Gr_1'-\Gr_0)^2/\Gr_1'']/\Gr_0^2, \quad \quad
[h_1''+(h_1'-h_0)^2/h_1'']=-[\Gk_1''+(\Gk_1'-\Gk_0)^2/\Gk_1'']/\Gk_0^2,
\eeq{8.4}
giving
\beqa 4\pi k_0\Imag(e^{2i\Go t_0}p^a P^s_\infty(-\Bk_0/k_0))/(\Go\Gr_0) 
& \leq &  -\frac{\Go(p^a)^2[\Gr_1''+(\Gr_1'-\Gr_0)^2/\Gr_1'']}{\Gr_0\Gk_0}\int_\GO\{\sin[(\Bk_0\cdot\Bx)+\Go t_0]\}^2\,d\Bx \nonum
& ~ & \quad-\frac{\Go(p^a)^2[\Gk_1''+(\Gk_1'-\Gk_0)^2/\Gk_1'']}{\Gk_0^2}\int_\GO\{\cos[(\Bk_0\cdot\Bx)+\Go t_0]\}^2\,d\Bx.
\eeqa{8.5}
where we have replaced $k_0^2$ with $\Go^2\Gr_0/\Gk_0$. This clearly then implies
\beq \frac{4\pi |P^s_\infty(-\Bk_0/k_0)|}{p^ak_0|\GO|}\leq \frac{[\Gr_1''+(\Gr_1'-\Gr_0)^2/\Gr_1'']}{\Gr_0}-\frac{[\Gk_1''+(\Gk_1'-\Gk_0)^2/\Gk_1'']}{\Gk_0},
\eeq{8.6}
in which $|\GO|$ is the volume of $\GO$ and $|P^s_\infty(-\Bk_0/k_0))|$ is the modulus of the backwards scattering amplitude $P^s_\infty(-\Bk_0/k_0))$. Note that both
terms on the right hand side of \eq{8.6} are non-negative because $\Gr_1''\geq 0$ and $\Gk_1'' \leq 0$. This bound implies that to ensure the backscattering is
small when $\Gr_1''$ and $\Gk_1''$ are small, one should match $\Gr_1'$ and $\Gk_1'$ to equal $\Gr_0$ and $\Gk_0$, respectively.
\section{Conclusion}
Perhaps the most important contribution of this paper is showing that Sommerfeld's radiation condition can be replaced by an appropriate ``constitutive law''
at infinity, akin to the perfectly matched layer (PML) technique in numerical analysis. The formulation of scattering as an appropriately defined $Y$-problem,
puts scattering under the umbrella of a wide class of problems and motives further investigation into the theory of $Y$-problems. It also raises the question
as to what other physical or mathematical problems can be reformulated as  $Y$-problems. It is interesting that the variational principles only give bounds on the
backward scattering amplitude rather than the desired forward scattering amplitude. We have no physical insight into why backscattering is subject to these bounds.
As shown in sections 7 and 8, some of the quantities first entering the variational principle are related to power dissipation and scattered power, and indeed this
was what motivated consideration of the quadratic form \eq{7.5}. However, surprisingly,
these power terms cancel out of the final variational principle. One wonders if the variational principles can be tweaked in some way to produce
bounds on the scattering amplitude in any direction.


\section*{Acknowledgements}
The author thanks the National Science Foundation for support through grant DMS-1211359. Also he thanks Steven Johnson and Owen Miller for stimulating discussions which
motivated this work.

\appendix
\section{Derivation of the variational inequality}
Here we derive the variational inequality \eq{5.17}. We have
\beq [\underline{\CF}_1^0+\underline{\CF}_2^0, \Imag(\underline{P}_\infty(\Bn)), \Real(\underline{P}_\infty(\Bn))]\in \CE,\quad
[\underline{\CG}_1^0+\underline{\CG}_2^0, \Real(\underline{V}_\infty(\Bn)), -\Imag(\underline{V}_\infty(\Bn)]\in \CJ.
\eeq{15.8}
Recall that $\Real (\CF^s_1)$ and $\Real (\CG^s_1)$ are prescribed as in \eq{5.9} that $\CF^s=\CF^s_1+\CF^s_2$ and $\CG^s=\CG^s_1+\CG^s_2$ are 
the associated solutions of the $Y$-problem. Since
\beq [\Real(\CF^s_1)+\Real(\CF^s_2), \Imag(P^s_\infty), \Real(P^s_\infty)] \in \CE, \quad [\Imag (\CF^s_1)+\Imag(\CF^s_2), -\Real(P^s_\infty), \Imag(P^s_\infty)]  \in \CE,
\eeq{15.10}
and
\beq
[\Real(\CG^s_1)+\Real(\CG^s_2), \Real(V^s_\infty), -\Imag(V^s_\infty)]\in \CJ, \quad [\Imag (\CG^s_1)+\Imag(\CG^s_2), \Imag(V^s_\infty), \Real(V^s_\infty)]\in \CJ,
\eeq{15.10a}
lie in orthogonal spaces, and since $\CV$ and $\CH$ are orthogonal, we deduce that
\beqa -(\Real(\CF^s_1),\Imag(\CG^s_1)) & = & \lang[\Real(\CF^s_2), \Imag(P^s_\infty), \Real(P^s_\infty)],[\Imag(\CG^s_2), \Imag(V^s_\infty), \Real(V^s_\infty]\rang \nonum
 & = & (\Real(\CF^s_2),\Imag(\CG^s_2))+\lang[0, \Imag(P^s_\infty), \Real(P^s_\infty)],[0, \Imag(V^s_\infty), \Real(V^s_\infty)]\rang, \nonum
(\Imag(\CF^s_1),\Real(\CG^s_1)) & = & -\lang[\Imag(\CF^s_2), -\Real(P^s_\infty), \Imag(P^s_\infty)],[\Real(\CG^s_2), \Real(V^s_\infty), -\Imag(V^s_\infty)]\rang \nonum
 & = & -(\Imag(\CF^s_2),\Real(\CG^s_2))+\lang[0, \Real(P^s_\infty), -\Imag(P^s_\infty)],[0, \Real(V^s_\infty), -\Imag(V^s_\infty)]\rang.
\eeqa{15.11}
Similarly \eq{15.8} with \eq{5.9} imply
\beqa -(\Real (\CF^s_1),\Imag(\CG^s_1))& = & 
\lang[\underline{\CF}_2^0, \Imag(\underline{P}_\infty), \Real(\underline{P}_\infty)],[\Imag(\CG^s_2), \Imag(V^s_\infty), \Real(V^s_\infty)]\rang \nonum
& = & (\underline{\CF}_2^0,\Imag(\CG^s_2))+\lang[0, \Imag(\underline{P}_\infty), \Real(\underline{P}_\infty)],[0,\Imag(V^s_\infty), \Real(V^s_\infty)]\rang, \nonum
 (\Imag(\CF^s_1),\Real (\CG^s_1))& = & 
-\lang[\Imag(\CF^s_2), -\Real(P^s_\infty), \Imag(P^s_\infty)],[\underline{\CG}_2^0, \Real(\underline{V}_\infty), -\Imag(\underline{V}_\infty)]\rang \nonum
& = & -(\Imag(\CF^s_2),\underline{\CG}_2^0)+\lang[0, \Real(P^s_\infty), -\Imag(P^s_\infty)],[0, \Real(\underline{V}_\infty), -\Imag(\underline{V}_\infty)]\rang.
\eeqa{15.12}
Now defining $\underline{\Bs}(\Bn)$ as in \eq{5.17a} we clearly have 
\beqa 0 &\leq & \int_{\GO}(\BE_2^0(\Bx)-\underline{\BE}_2^0(\Bx))\cdot\BCL_1(\BE_2^0(\Bx)-\underline{\BE}_2^0(\Bx))\,d\Bx
+\int_{|\Bn|=1} (\Bs(\Bn)-\underline{\Bs}(\Bn))\cdot\BCL_3(\Bs(\Bn)-\underline{\Bs}(\Bn))\,dS \nonum
&=& \int_{\GO}\underline{\BE}_2^0(\Bx)\cdot\BCL_1\underline{\BE}_2^0(\Bx)\,d\Bx+\int_{|\Bn|=1}\underline{\Bs}(\Bn)\cdot\BCL_3\underline{\Bs}(\Bn)\,dS
-2(\underline{\BE}_2^0,\BJ_2^0)+(\BE_2^0,\BJ_2^0) \nonum
&~&\quad -2\int_{|\Bn|=1} \underline{\Bs}(\Bn)\cdot\Bt(\Bn)\,dS
+\int_{|\Bn|=1} \Bs(\Bn)\cdot\Bt(\Bn)\,dS,
\eeqa{15.14}
where $\Bs(\Bn)$ and $\Bt(\Bn)$ are defined in \eq{5.3}, and we have used the constitutive laws \eq{5.2} and \eq{5.3}, and the identities \eq{5.2b} and \eq{5.7b}. Since
\beqa 
(\BE_2^0,\BJ_2^0)& = & -(\Real(\CG^s_2),\Imag (\CF^s_2))+(\Real(\CF^s_2), \Imag(\CG^s_2)), \nonum
\int_{|\Bn|=1} \Bs(\Bn)\cdot\Bt(\Bn)\,dS & = & \lang [0,\Real(V^s_\infty), -\Imag(V^s_\infty)],[0,\Real(P^s_\infty),-\Imag(P^s_\infty) ]\rang \nonum
&~& +\lang [0, \Imag(P^s_\infty), \Real(P^s_\infty)],[0,\Imag(V^s_\infty),\Real(V^s_\infty)]\rang, \nonum
(\underline{\BE}_2^0,\BJ_2^0)& = & -(\underline{\CG}_2^0,\Imag (\CF^s_2))+(\underline{\CF}_2^0, \Imag(\CG^s_2)), \nonum
\int_{|\Bn|=1} \underline{\Bs}(\Bn)\cdot\Bt(\Bn)\,dS & = & \lang [0,\Real(\underline{V}_\infty), -\Imag(\underline{V}_\infty)],[0,\Real(P^s_\infty),-\Imag(P^s_\infty) ]\rang \nonum
&~& +\lang [0, \Imag(\underline{P}_\infty), \Real(\underline{P}_\infty)],[0,\Imag(V^s_\infty),\Real(V^s_\infty)]\rang, 
\eeqa{15.15}
the identities \eq{15.12} imply
\beqa (\BE_2^0,\BJ_2^0)+\int_{|\Bn|=1} \Bs(\Bn)\cdot\Bt(\Bn)\,dS & = & (\Imag(\CF^s_1),\Real(\CG^s_1))-(\Real(\CF^s_1),\Imag(\CG^s_1)), \nonum
(\underline{\BE}_2^0,\BJ_2^0)+\int_{|\Bn|=1} \underline{\Bs}(\Bn)\cdot\Bt(\Bn)\,dS & = &  (\Imag(\CF^s_1),\Real (\CG^s_1))-(\Real (\CF^s_1),\Imag(\CG^s_1)).
\eeqa{15.16}
Substituting these in \eq{15.14} gives the variational inequality \eq{5.17}.
\section{Connection with Optical Theorems}
Here we show that the expression \eq{6.10f} for the extinction power can be connected to other expressions for $W$, that are generally
known as optical theorems. From \eq{6.10f} it follows that
\beqa 2W & = & \Imag(\CF^a+\CF_1^s,\Gc\BZ_1(\CF^a+\CF_1^s))-\Imag(\CF_1^s,\CG_1^s)\nonum
& = & \Imag(\CF^a,\Gc\BZ_1\CF^a)+\Imag(\CF_1^s,\Gc\BZ_1\CF^a)+\Imag(\CF^a,\BZ_1\CF_1^s)+\Imag(\CF_1^s,\BZ_1\CF_1^s)\nonum
& ~& -\Imag(\CF_1^s,\BZ_1\CF_1^s+(\BZ_1-\BZ_0)\Gc\CF^a)\nonum
& = & \Imag(\CF^a,\Gc\BZ_1\CF^a)+\Imag(\CF^a,\BZ_1\CF_1^s)+\Imag(\CF_1^s,\BZ_0\Gc\CF^a).
\eeqa{16.10}
Since $\BZ_0$ is real we also have
\beqa 0 & = & \Imag(\CF^a+\CF_1^s,\Gc\BZ_0(\CF^a+\CF_1^s)) \nonum
& = &\Imag(\CF^a,\Gc\BZ_0\CF^a)+\Imag(\CF_1^s,\Gc\BZ_0\CF^a)+\Imag(\CF^a,\BZ_0\CF_1^s)+\Imag(\CF_1^s,\BZ_0\CF_1^s) \nonum
& = & \Imag(\CF_1^s,\Gc\BZ_0\CF^a)+\Imag(\CF^a,\BZ_0\CF_1^s).
\eeqa{16.11}
Substituting this back in \eq{16.10}, and again using the fact that
$\Imag(\CF^a,\Gc\BZ_0\CF^a)=0$, gives
\beqa 2W & = & \Imag(\CF^a,\Gc(\BZ_1-\BZ_0)\CF^a)+\Imag(\CF^a,\BZ_1\CF_1^s)+ \Imag(\CF^a,\BZ_0\CF_1^s)\nonum
& = & \Imag(\CF^a,\Gc(\BZ_1-\BZ_0)(\CF^a+\CF_1^s))\nonum
& = & \Imag(\CF^a,\Gc(\BZ_1-\BZ_0)(\CF^a+\CF^s)).
\eeqa{16.12}
This is analogous to the form of the optical theorem given in \cite{Lytle:2005:GOT,Hashemi:2012:DBP,Miller:2015:FLO}. We can further reduce it to
\beqa
2W & = & \Imag(\CF^a,(\BZ-\BZ_0)(\CF^a+\CF^s)) \nonum
& = & \Imag(\CF^a,\CG^s-\BZ_0\CF^s) \nonum
& = & \Imag\int_{B_{r}}\overline{p^a}{\CF^a}'\cdot(\CG^s-\BZ_0\CF^s)\,d\Bx,
\eeqa{16.13}
where ${\CF^a}'=\overline{\CF^a/p^a}$. Thus using the results of Section \sect{farfield}, and making the substitution $\Bk_0'=\Bk_0$ we get
\beqa W & = & 2k_0\pi\Imag[\overline{p^a} P^s_\infty(\Bk_0/k_0)]/(\Go\Gr_0),
\eeqa{16.14}
which is the well-known form of the optical theorem \cite{Newton:1976:OTB,Dassios:2000:LFS}
for acoustic scattering.

\bibliographystyle{../siamplain}
\bibliography{/home/milton/tcbook,/home/milton/newref}
\end{document}